\documentclass{article}

\PassOptionsToPackage{numbers, compress}{natbib}
\usepackage[nonatbib, final]{neurips_2024}

\usepackage[utf8]{inputenc}
\usepackage[T1]{fontenc}
\usepackage{xurl}
\usepackage[colorlinks, allcolors=mydarkblue]{hyperref}
\usepackage{booktabs}
\usepackage{amsfonts}
\usepackage{nicefrac}
\usepackage{microtype}
\usepackage{xcolor}
\usepackage{graphicx}
\usepackage{tabularx}
\usepackage{xltabular}
\usepackage{enumitem}
\usepackage{multirow}
\usepackage{array}
\usepackage{apacite}
\usepackage{cleveref}
\let\citep\shortcite
\let\citet\shortciteA
\let\citealp\shortciteNP
\AtBeginDocument{\urlstyle{APACsame}}
\definecolor{mydarkblue}{rgb}{0,0.08,0.45}

\title{Embedded Assessments for Frontier AI}

\author{%
\parbox{\linewidth}{\centering%
\mbox{Jacob Charnock\thanks{Work completed as a seasonal fellow at GovAI. Corresponding author: \url{jakecharnock25@gmail.com}.}\hspace{1.15ex}$^{1}$\enskip} %
\mbox{Sophie Williams\hspace{0.2ex}$^{1}$\enskip} %
\mbox{Zaheed Kara\hspace{0.2ex}$^{1}$\enskip} \\ \rule[-1mm]{0pt}{7mm}
\mbox{Markus Anderljung\hspace{0.2ex}$^{1}$\enskip} %
\mbox{Alejandro Tlaie Boria\hspace{0.2ex}$^{2}$\enskip} %
\mbox{Stephen Casper\hspace{0.2ex}$^{3}$\enskip} \\ \rule[-1mm]{0pt}{7mm}
\mbox{Anka Reuel\hspace{0.2ex}$^{4}$\enskip} %
\mbox{Jonas Freund\hspace{0.2ex}$^{1}$\enskip}\\%
}
\\\\%
$^{1}$GovAI \enskip $^{2}$Pour Demain \enskip $^{3}$Harvard Kennedy School \enskip $^{4}$Stanford University}

\begin{document}
\maketitle
\setcounter{footnote}{0}

\begin{abstract}
Third-party evaluations for frontier AI have mostly tested models through external interfaces before deployment. But the risks from frontier AI models depend on how their developers use and govern them internally. Recently, CEOs of frontier AI companies have committed to hosting embedded assessments. These assessments would give independent evaluators employee-like access to a developer's internal systems, staff, and documentation. First, we argue that this can enable deeper and more flexible assessments of risks that depend on internal systems and practices, while providing access under stronger security controls. Then, we examine seven design questions about scope, information gathering, duration, timing, terms of engagement, disclosure, and escalation. We recommend that frontier AI developers begin hosting embedded assessments now, covering at least three areas central to managing risks from internal AI use: internal agent monitoring, internal agent security controls and permissions, and model alignment. To enable meaningful third-party scrutiny, assessments should be continuous, evaluators should publish detailed reports at least quarterly, and clear escalation mechanisms should be established. These recommendations are intended as a starting point, with further steps needed to realize the full potential of embedded assessments.
\end{abstract}
\vspace{1.6em}

\newpage
\tableofcontents
\newpage

\section*{Executive Summary}
\addcontentsline{toc}{section}{Executive Summary}

Third-party evaluations for frontier AI have largely involved testing models through external interfaces before deployment, providing limited scrutiny of risks that arise from developers' internal systems and practices. Recent incidents illustrate the consequences of this gap. Most notably, OpenAI models circumvented restrictions on internet access during internal testing and compromised Hugging Face's production infrastructure.

This paper makes the case for embedded assessments, a form of third-party evaluation in which independent evaluators are given access to a frontier AI developer's internal systems, staff, and documentation. It also discusses practical questions about how they could be conducted, and makes recommendations for how they could build on recent pilots and develop into a more established industry practice.

\textbf{What are embedded assessments? (\Cref{sec:1})}

Embedded assessments involve providing independent evaluators with employee-like access to a frontier AI developer's internal systems, staff, and documentation, typically while working within the developer's physical offices. This enables evaluators to assess risks that depend on the developer's live infrastructure and internal practices, which are difficult (or even impossible) to evaluate directly and securely from externally shareable artifacts alone. Similar forms of onsite assessments are already used in other high-stakes industries, including nuclear power, banking, and food and drug manufacturing.

In March 2026, Model Evaluation and Threat Research (METR) conducted the first embedded assessment of a frontier AI developer's internal systems. It assessed Anthropic's internal agent-monitoring systems and successfully identified several previously unknown vulnerabilities. In July and August 2026, METR and Redwood Research conducted an embedded assessment at OpenAI. They investigated the behavior of OpenAI's internal agents involved in an incident in which the models hacked another company's production infrastructure.

However, both these assessments were limited in several important ways, reflecting their purposes as an initial pilot and a targeted incident investigation, respectively. They involved only a small number of evaluators, ran for short periods of time, and assessed a narrow range of internal activity. Since then, the CEOs of Anthropic and OpenAI have committed to hosting embedded assessments, with Anthropic's CEO committing to an ongoing arrangement.

\textbf{The case for embedded assessments (\Cref{sec:2})}

Embedded assessments offer five main benefits:

\begin{itemize}[leftmargin=2em]
    \item They enable evaluators to access sensitive systems under stronger, physical security controls than would be possible in remote assessments.
    \item They allow evaluators to better assess risks that depend on the developer's internal systems and practices, including risks from internal AI use.
    \item They can make assessments deeper and more flexible by giving evaluators richer contextual information and making it easier to investigate unexpected findings.
    \item They can help identify and remediate vulnerabilities before they lead to incidents.
    \item They can accelerate post-incident analysis and response by establishing access arrangements in advance and improving evaluators' familiarity with the developer's systems.
\end{itemize}

Embedded assessments also present several challenges, but these can be addressed through careful design choices, as in other industries:

\begin{itemize}[leftmargin=2em]
    \item They may pose greater intellectual property risks than remote evaluations because onsite evaluators are likely to encounter more sensitive information, sometimes incidentally. Possible mitigations include supervising evaluator activity, physically separating evaluators from operations unrelated to the assessment, limiting what information evaluators can take offsite, and using personnel vetting and contractual cooling-off periods.
    \item Developers may be concerned that embedded assessments will uncover improprieties that create reputational or legal risks. These risks can be managed in part through clear procedures for publishing findings, redacting sensitive information, and handling legally compelled disclosure.
    \item Evaluators and developers need to dedicate time and resources to the assessment, which may slow or displace other work. Options for reducing these costs include using contractual templates to avoid having to draft agreements from scratch and appointing dedicated company liaisons to coordinate the review.
\end{itemize}

\textbf{Design choices for conducting embedded assessments (\Cref{sec:3})}

We identify seven design choices for embedded assessments and discuss options for each. The first concerns which systems and practices evaluators could examine. We focus on areas that are particularly important for managing risks from internal AI use. Embedded assessments should also extend to other areas (see the Appendix for potential applications), but these are beyond the scope of this paper. The other six design choices apply across embedded assessments more generally.

\renewcommand{\theHtable}{uncaptioned.summary}
\renewcommand{\arraystretch}{1.2}
\begin{xltabular}{\linewidth}{%
        >{\small\raggedright\arraybackslash}p{\dimexpr0.35\linewidth-2\tabcolsep\relax}%
        >{\small\raggedright\arraybackslash}X}
    \toprule
        \textbf{Question} &
        \textbf{Options}\\
    \midrule
    \endfirsthead
    \toprule
        \textbf{Question} &
        \textbf{Options}\\
    \midrule
    \endhead
    \bottomrule
    \endfoot
    \bottomrule
    \endlastfoot
        What systems and practices could evaluators examine? &
        \vspace{-0.7em}\begin{itemize}[leftmargin=*, nosep]
\item Internal agent monitoring (assessing whether monitors adequately cover consequential agent activity and reliably flag harmful behavior, and whether resulting flags are escalated and reviewed appropriately)
\item Internal agent security controls and permissions (assessing the restrictions placed on internal AI agents to prevent harmful behavior, and whether internal agents can bypass them)
\item Model alignment (assessing whether models exhibit concerning behaviors, such as deception or reward hacking, throughout their entire lifecycle, and whether current alignment techniques are likely to remain effective in the future)
\end{itemize}\vspace{-0.2em}\\
        How could evaluators gather information? &
        \vspace{-0.7em}\begin{itemize}[leftmargin=*, nosep]
\item Infrastructure access (e.g. monitoring systems, sandboxes, training environments, internal models)
\item Personnel access (e.g. interviews, questionnaires, demonstrations, meetings, messaging channels)
\item Documentation access (e.g. assessment results, agent logs, incident reports)
\end{itemize}\vspace{-0.2em}\\
        How long could assessments last? &
        \vspace{-0.7em}\begin{itemize}[leftmargin=*, nosep]
\item Fixed duration
\item Flexible duration
\item Continuous
\end{itemize}\vspace{-0.2em}\\
        When could assessments take place? &
        \vspace{-0.7em}\begin{itemize}[leftmargin=*, nosep]
\item Ad hoc (at the developer's discretion)
\item Periodic
\item Event-based
\item Unannounced (at the evaluator's discretion)
\end{itemize}\vspace{-0.2em}\\
        What terms could govern the assessment? &
        \vspace{-0.7em}\begin{itemize}[leftmargin=*, nosep]
\item Access rights
\item Security and information-handling requirements
\item Evaluator independence
\item Legal protections
\end{itemize}\vspace{-0.2em}\\
        How could findings be publicly disclosed? &
        \vspace{-0.7em}\begin{itemize}[leftmargin=*, nosep]
\item No information about the review
\item A statement that the review occurred
\item A summary report
\item A detailed report
\end{itemize}\vspace{-0.2em}\\
        What escalation procedures could be put in place? &
        \vspace{-0.7em}\begin{itemize}[leftmargin=*, nosep]
\item No formal escalation procedure
\item Escalation to a designated senior employee responsible for safety at the developer
\item Escalation to an independent oversight body (e.g. the developer's board of directors)
\item Escalation to a government body
\end{itemize}\vspace{-0.2em}\\
\end{xltabular}
\addtocounter{table}{-1}
\renewcommand{\theHtable}{\arabic{table}}
\newpage

\textbf{Our recommendations (\Cref{sec:4})}

We recommend that, at a minimum, embedded assessments cover three areas of internal AI risk management: internal agent monitoring, internal agent security controls and permissions, and model alignment. Assessments of internal agent monitoring would examine whether monitors adequately cover consequential agent activity and reliably flag harmful behavior, as well as whether resulting flags are escalated and reviewed appropriately. Assessments of security controls and permissions would examine what restrictions are placed on internal AI agents and whether agents can bypass those restrictions. Assessments of model alignment would involve conducting evaluations to assess whether models remain aligned throughout their entire lifecycle, and whether current alignment techniques are likely to remain effective for more capable models in the future. These three areas are central to managing risks from internal AI use, appear tractable to assess in the near term, and are difficult to scrutinize meaningfully without the employee-like access that embedded assessments provide.

To carry out such assessments, evaluators should receive access to relevant infrastructure, personnel, and documentation, with access parity to internal employees conducting similar risk assessments or senior alignment researchers, so they can consider all relevant information needed to reach thorough, independent conclusions. Evaluators should receive access to any additional information they request unless the developer has a specific reason to deny it, and, in such cases, the developer should share those reasons with the evaluator.

Given the rapid pace of progress in frontier AI capabilities, embedded assessments should operate continuously, allowing evaluators to track changes, investigate emerging concerns, and test whether corrective measures remain effective. Serious incidents or material changes should prompt targeted investigations. Where continuous assessment is not yet feasible, flexible-duration reviews conducted periodically, with event-triggered follow-up, should serve as an interim approach rather than an equivalent alternative.

Pre-agreed terms should govern the assessment, including access rights, security and information-handling requirements, evaluator independence, and legal protections. Agreeing these terms in advance can help ensure evaluators receive sufficient access and independence to conduct a credible assessment while allowing developers to manage security, intellectual property, and legal risks.

Evaluators should publish detailed public reports at least quarterly, explaining the scope of the assessment, their findings and supporting evidence, key limitations and uncertainties, and any recommendations that have not yet been addressed. Before publication, the report should be shared with the developer, who should have a narrowly defined right to redact sensitive information. The evaluator should retain editorial control to ensure that findings are not redacted simply because they are unfavorable and should indicate whether any redactions are material to the report's conclusions.

The developer should designate a senior employee responsible for safety and make them formally accountable for reviewing the findings and recommendations and overseeing the developer's response. Serious or unresolved concerns should be escalated to the board of directors or another oversight body (e.g. OpenAI's Safety and Security Committee and Anthropic's Long-Term Benefit Trust). The developer should publish a follow-up report setting out which recommendations it intends to adopt and a plan for doing so, and periodically report its progress.

While this paper focuses on risk management for internal AI use, embedded assessments should go beyond this, allowing evaluators to conduct investigations across a wider range of areas and surface more ``unknown unknowns''. This could extend independent scrutiny to other important aspects of frontier AI development and deployment, including security mitigations protecting sensitive assets (e.g. model weights) and compliance with potential future pacing commitments.

As embedded assessments mature, they should also provide stronger external accountability. Particularly serious or unresolved concerns should be escalated to an appropriate government body in cases where other governance mechanisms have failed to address them. It will also become increasingly important to ensure that evaluator teams collectively have the expertise needed to assess increasingly broad and technically demanding areas, including by involving multiple evaluation organizations where necessary.
\newpage

\section{Introduction}\label{sec:1}

Third-party evaluations are becoming an increasingly important part of frontier AI risk management \citep{fmf_2025, metr_2025}. Many frontier AI developers state in their safety frameworks that they use third-party evaluations where appropriate.\footnote{For example: Anthropic's Responsible Scaling Policy states that capability and alignment evaluations will be ``conducted internally and by external parties as appropriate'' \citep{anthropic_2026i}; OpenAI's Preparedness Framework states that ``when available and feasible, OpenAI will work with third-parties to independently evaluate models'' \citep{openai_2025}; and Google DeepMind's Frontier Safety Framework states that it incorporates ``external evaluations'' into its risk assessments ``as appropriate'' \citep{googledeepmind_2026}.} Third-party evaluations also appear in regulatory requirements, including the EU AI Act's General-Purpose AI Code of Practice \citep{europeancommission_2025} and Illinois' SB 315 \citep{illinois_2026}.\footnote{Appendix 3.5 of the Safety and Security chapter of the EU's General-Purpose AI Code of Practice requires signatories to ``ensure that adequately qualified independent external evaluators conduct model evaluations'' pursuant to Appendix 3. Illinois' SB 315 § 10(a)(5) requires large frontier developers to publish a frontier AI framework describing, among other things, their approach to ``using third parties to assess the potential for catastrophic risks and the effectiveness of mitigations of catastrophic risks''. In § 10(d), it additionally requires large frontier developers to ``annually retain a third party to perform an independent audit of compliance with the requirements of this Section''.}

To date, however, most third-party evaluations for frontier AI have not provided evaluators with access to developers' live internal systems and operating environments. Instead, they have typically involved testing model checkpoints through an API (e.g. \citealp{anthropic_2026a, metr_2026b, metr_2026a, openai_2026a, securebio_2026}).\footnote{This includes assessments of pre-release model checkpoints (e.g. \citealp{anthropic_2026a, openai_2026a}), helpful-only or rail-free model variants \citep{metr_2026a, securebio_2026}, and internal state-of-the-art models \citep{metr_2026b}.} This limits evaluators' ability to assess how models interact with the systems and processes in which they are developed and used. This can make it difficult for third parties to assess whether developers' internal safeguards and practices are sufficient to manage the risks posed by frontier AI. Recent incidents illustrate the consequences of this gap: weaknesses in developers' internal monitoring and containment have allowed AI agents to evade oversight, escape sandboxes, and reach external infrastructure before these systems had received meaningful independent scrutiny \citep{anthropic_2026b, openai_2026c, openai_2026b}.

Embedded assessments provide independent evaluators with access to relevant internal systems, staff, and nonpublic information comparable to that of internal employees who conduct similar risk assessments or to senior alignment researchers. As a result, embedded evaluators have greater visibility into the technical environment and the broader sociotechnical context in which systems are developed and used by the developer. This enables relatively secure access to sensitive systems, allows evaluators to assess risks that existing evaluations do not meaningfully address, and increases the depth and flexibility of assessments. It can also help identify vulnerabilities before they lead to incidents, and accelerate incident analysis and response. In theory, embedded evaluators could work onsite or remotely (with access similar to that of remote employees). However, we focus on onsite assessments because we expect them to offer additional benefits, for example by providing evaluators with richer organizational context, more immediate interaction with staff, and stronger controls for accessing sensitive systems. Unless otherwise specified, the use of the term embedded assessments in this paper refers to onsite arrangements.

Despite their potential, embedded assessments remain underdeveloped as an approach to frontier AI oversight. Onsite evaluations are common in other domains such as nuclear power (\href{https://www.ecfr.gov/current/title-10/chapter-I/part-50/section-50.70}{10 C.F.R. § 50.70}), banking (\href{https://www.law.cornell.edu/uscode/text/12/481}{12 U.S.C. § 481}), meat production (\href{https://www.law.cornell.edu/uscode/text/21/603}{21 U.S.C. § 603}), food and drug manufacturing \citep{fda_2025}, food hygiene \citep{fsa_2023}, and international nuclear safeguards \citep{iaea_nd}. Moreover, several papers have argued that frontier AI risk management could draw on more established industries that use onsite reviews \citep{brundage_2026, contibrown_2026, wills_2025}. However, these papers do not set out what embedded assessments for frontier AI could or should involve in practice.

In March 2026, METR piloted the first embedded assessment for frontier AI, demonstrating that such engagements are both feasible and valuable \citep{metr_2026c}. During the assessment, one METR staff member spent three weeks red-teaming a subset of Anthropic's internal agent monitoring systems. In July and August 2026, METR and Redwood Research conducted a second such engagement. They worked onsite at OpenAI for six days to investigate agent behavior during an incident, in which internal models escaped their sandbox and compromised Hugging Face's production infrastructure \citep{metr_2026d}. However, these early engagements had important limitations. For example, the OpenAI-Hugging Face investigation was constrained by limited access and capacity: evaluators could not query the model involved or directly access relevant infrastructure data, and the scale of the incident meant that substantial analysis had to be delegated to often-unreliable AI agents \citep{metr_2026d}.

Since then, Anthropic's CEO, Dario Amodei, has said publicly that Anthropic is ``unilaterally committing'' to implementing embedded evaluators, as they are key to verifying that frontier AI developers are following their stated safety practices \citep{amodei_2026}. OpenAI's CEO, Sam Altman, subsequently committed to providing independent evaluators with employee-like access as well \citep{altman_2026}. Other frontier AI company CEOs, including xAI's Elon Musk and Microsoft's Satya Nadella, have also signaled general support for the approach \citep{musk_2026, nadella_2026}. Additionally, the AI Evaluator Forum has endorsed embedded assessments and outlined some minimum conditions for ensuring their credibility \citep{aef_2026}. Recent proposals have begun to identify technical priorities and pilot methods for embedded assessments \citep{openmined_2026, transluce_2026}. This paper builds on these proposals by examining the broader design choices that developers and evaluators need to resolve.

Embedded assessments could be used to assess many areas of frontier AI developers' activities, including their security mitigations for protecting sensitive assets or even compliance with potential future pacing commitments (see the Appendix). However, we focus on the management of risks arising from frontier AI developers' internal use of AI. By internal AI use, we mean frontier AI developers' use of AI models in their own research, development, and operations. We take this focus for three reasons. First, risks from internal AI use are high priority and can materialize unexpectedly. Frontier AI developers are increasingly using their most capable models for safety-critical tasks and to automate parts of their own AI R\&D \citep{acharya_2025, stix_2025}. Recent incidents have illustrated the potential risks of internal AI use, with internal models evading monitoring, escaping sandboxes, and reaching the open internet due to weaknesses in internal controls \citep{anthropic_2026b, openai_2026d}.\footnote{For example, in July 2026, models internal to OpenAI exploited a vulnerability in their sandbox environment, evaded internal monitoring, and compromised Hugging Face's production infrastructure during internal testing \citep{openai_2026d}. Anthropic subsequently disclosed three incidents in which internal models undergoing cybersecurity evaluations accessed the internet through misconfigured evaluation environments and gained unauthorized access to the production systems of three organizations \citep{anthropic_2026b}. In August 2026, it was reported that a misconfigured evaluation environment allowed one of Meta's models to access the internet and compromise another company's systems during cybersecurity testing \citep{npr_2026}.} Second, internal AI use is already a tractable focus for embedded assessments. Pilot embedded assessments at Anthropic and OpenAI suggest that independent review of internal AI use is feasible and valuable, though as pilots they were limited in scope and did not cover all safety-critical aspects of internal AI use. Third, risks from internal AI use are particularly well-suited to embedded assessments because assessing them requires employee-like access. This allows evaluators to understand how models are used and overseen in practice, including their task assignments, access, permissions, monitoring, and other safeguards. Without this access, evaluators may be unable to assess relevant systems directly or obtain enough context to assess these risks reliably.

This paper is primarily intended for frontier AI developers, evaluation organizations, and policymakers. For developers and evaluators, we provide a framework to inform the design of embedded assessments. For developers, we also set out ways to address the challenges such assessments may pose. We also expect the paper to be of interest to policymakers considering how embedded assessments could advance the state of the art in third-party scrutiny of frontier AI risks.

Against this background, this paper makes three contributions. First, we discuss the benefits and challenges of embedded assessments (\Cref{sec:2}). Second, we explore how embedded assessments could be designed in practice (\Cref{sec:3}). Finally, we set out recommendations for embedded assessments now, as well as some high-priority steps to broaden their scope and strengthen their role in AI oversight (\Cref{sec:4}).

\section{The case for embedded assessments}\label{sec:2}

In this section, we make the case for embedded assessments. We discuss the main benefits (\Cref{sec:2.1}), as well as the main challenges and how these could be addressed (\Cref{sec:2.2}).

\subsection{Benefits}\label{sec:2.1}

Embedded assessments offer five main benefits, which apply in different ways to frontier AI developers, evaluators, and external stakeholders. For developers and evaluators, they can enable deeper and more flexible assessments under stronger security controls. For developers in particular, they can help identify vulnerabilities earlier and support faster incident response, potentially reducing costs. For external stakeholders, they can provide a stronger basis for judging whether frontier AI risks are being managed effectively and, ultimately, help reduce wider harms. These benefits are discussed in more detail below.

\textbf{Embedded assessments can enable deeper access under stronger security controls.} Comprehensively assessing a developer's risk management practices may require access to the developer's internal systems and other sensitive information. This is particularly true for some types of risks, such as those arising from a developer's internal use of its models. While in theory this is possible to do offsite, remote evaluations often require developers to expose information, such as model-access credentials, over the internet. This creates risks if an evaluator's security practices are weak or its network is compromised.\footnote{Stolen credentials are one of the most common ways attackers first break into an organization's systems, and leaked keys can stay valid for months \citep{cisa_2024, mandiant_2025, verizon_2025}.} Compromised access could potentially enable attackers to exfiltrate model weights, remove safety controls, or insert backdoors \citep{hurel_2026}. These risks may be especially significant for grey-box and white-box interfaces that provide access to model internals such as logits, activations, fine-tuning capabilities, or model weights. Embedded assessments can mitigate some of these risks through strong physical and organizational safeguards (e.g. company-managed devices, escort protocols, and secure review rooms), as is standard practice in other industries \citep{aonhewitt_2010, casper_2024, dod_2012, ecb_2018, iaea_1972, nist_2020}.\footnote{In European Central Bank onsite inspections, the inspected bank must provide the inspection team with secured offices and individual workstations in a lockable room, and inspectors receive read-only access to relevant systems \citep[§3.3.3]{ecb_2018}. International Atomic Energy Agency safeguards inspectors are limited to agreed ``strategic points'' within declared facilities, and the inspected state may have its representatives accompany them \citep{iaea_1972}. In US facilities accredited for sensitive compartmented information, visitors surrender personal electronic devices and uncleared personnel remain under constant escort by cleared staff \citep{dod_2012}. US federal security baselines similarly restrict non-organizationally owned systems, including personally owned devices, from processing, storing, or transmitting organizational information \citep[AC-20(3)]{nist_2020}. In mergers and acquisitions, competitively sensitive information is shared before a deal closes only inside ``clean rooms'', isolated physical or digital environments where access is logged and material cannot be removed \citep{aonhewitt_2010}.} This is beneficial to developers because it allows for rigorous assessment while reducing the risks that sensitive information is leaked. Enabling deeper access to sensitive information is the main benefit of embedded assessments. This builds on earlier proposals to give independent auditors privileged, secure access to confidential information so they can verify developers' claims \citep{brundage_2020}. Much of the value discussed below depends on the deeper access that these arrangements enable.

\textbf{Embedded assessments can help evaluators assess risks that may be difficult to assess remotely.} Some risks -- including those arising from internal AI use -- depend on a developer's internal infrastructure and practices. Relevant internal features can include task assignments, permissions, and monitoring systems \citep{stix_2025}. For example, in a pilot embedded assessment, METR noted a flaw in Anthropic's live attribution logic, whereby changing a single environment variable caused internal agent traffic to be misclassified, substantially reducing the likelihood that it would be monitored. METR described this flaw as ``highly non-obvious'' without specific knowledge of Anthropic's infrastructure \citep{metr_2026b, metr_2026c}. Although developers could reproduce some of these features for remote evaluators, it may be difficult to verify that the reproduced environment accurately reflects the systems operating inside the company. A frontier AI developer's internal systems evolve continuously as models are fine-tuned, integrated into new workflows, and given different tools and permissions. Training and operational environments are also updated frequently, creating similar challenges. Keeping remote evaluation environments up-to-date could therefore be more prone to error and potentially even more cumbersome for developers \citep{kwon_2026}.

\textbf{Embedded assessments can increase the depth and flexibility of assessments.} Onsite assessments can give evaluators richer contextual information than APIs, model checkpoints, or documentation alone. This context can help them develop a deeper understanding of the risks \citep{casper_2024}. For example, evaluators could interview staff who build and operate safety systems, attend relevant meetings, and review internal communications and documentation. This could reveal assumptions, implementation details, or potential failure modes that would otherwise be difficult to identify. Onsite access could also make it easier to follow up as new questions emerge during an assessment. Evaluators could seek additional information from staff or request further evidence and testing where the initial information is incomplete or raises new lines of inquiry. During their evaluation at OpenAI, METR and Redwood Research asked questions to clarify how the datasets they were analyzing had been constructed and requested a more comprehensive set of transcripts to address some limitations in the information initially provided \citep{metr_2026d}. While offsite evaluators can also request additional information and interview staff, embedded arrangements could reduce the friction of these exchanges, particularly where the requested information is sensitive. Embedded evaluators may also be better placed to identify omissions or inconsistencies in information provided by the developer as they can compare it with what they observe in practice. Embedded assessments could therefore provide greater coverage, assurance, and depth, even for risks that could in principle be assessed remotely.

\textbf{Embedded assessments could help identify and remediate vulnerabilities before they lead to incidents.} Recent incidents suggest that weaknesses and gaps in internal monitoring and security systems exist. For example, the OpenAI-Hugging Face incident exposed gaps in OpenAI's cyber protections and monitoring during internal testing \citep{openai_2026d}. Anthropic has likewise reported cyber incidents that arose during its own testing that initially went undetected \citep{anthropic_2026b}. Embedded assessments can help evaluators uncover these kinds of vulnerabilities in internal systems and processes. For example, in its pilot embedded assessment, METR was able to uncover novel vulnerabilities in Anthropic's internal monitoring \citep{metr_2026c}. Access to the relevant systems could also allow evaluators to test proposed fixes under realistic operational conditions. This reduces the risk of incidents and, in turn, the risk of harm to users, other organizations, and the wider public. By reducing the frequency and severity of incidents, embedded assessments could also reduce costs for developers. For example, OpenAI noted that as a result of the Hugging Face incident it ``temporarily slowed model development'' and ``incurred great cost and delays to frontier research'' \citep{openai_2026e}.

\textbf{Embedded assessments could help facilitate and accelerate incident analysis and response}. Independent investigations of AI incidents can require access to specialized infrastructure, tools, or sensitive information, which embedded assessments can provide. For example, investigators may require access to agent transcripts, intermediate model checkpoints, relevant operational environments, and sufficient inference resources \citep{metr_2026e, safe_2026}. Having evaluators embedded before an incident occurs could also enable investigations to take place more quickly. One reason for this is that evaluators may already be familiar with the relevant systems and infrastructure for analysis. The legal agreements and working arrangements needed to access relevant systems may also already be in place. Earlier assessments may also have surfaced and resolved problems with access, permissions, and coordination before they impede an investigation \citep{nist_2025}. Faster incident analysis and response can help limit the amount of resulting harm and reduce costs for developers, including potential liability and the operational costs associated with pausing activities.

\subsection{Challenges}\label{sec:2.2}

At the same time, embedded assessments may pose some challenges. Below, we set out three key challenges and suggest ways to address them.

\textbf{Developers may be concerned that embedded assessments can create intellectual property risks.} By placing evaluators inside the developer, and giving them access to internal systems and staff, embedded assessments increase both the amount and sensitivity of proprietary information that evaluators encounter. This can include both explicit and tacit knowledge that has the potential to benefit competitors or create misuse risks if leaked \citep{hurel_2026}.\footnote{Explicit knowledge includes detailed information about internal AI-use workflows, model-development processes, or R\&D strategy. Tacit knowledge includes how engineers approach security problems, improve agent performance, or understand weaknesses in existing systems.} During an embedded assessment, evaluators may also encounter sensitive information incidentally, so it is harder for the developer to control exactly what they learn. For example, evaluators might overhear conversations or observe systems outside the intended scope of the review. Employees may also disclose more to onsite reviewers through informal conversations than they would through formal, documented channels, potentially including concerns they have not raised with management \citep{homewood_2025}.

These risks can be reduced by limiting both what evaluators can access and what information evaluators can take offsite. Access can be restricted through least-privilege permissions (e.g. by ensuring embedded evaluators' access is scoped to that of relevant internal staff) and supervision of evaluator activity (\citealp{ecb_2018, nist_2020}; \href{https://www.ecfr.gov/current/title-10/chapter-I/part-73/section-73.55}{10 C.F.R. § 73.55}).\footnote{Federal security standards require limiting users to the access necessary for their assigned tasks \citep{nist_2020}, nuclear facilities must escort all visitors within protected areas (\href{https://www.ecfr.gov/current/title-10/chapter-I/part-73/section-73.55}{10 C.F.R. § 73.55}), and ECB banking inspections are carried out on the basis of a predefined scope, timeline, and set of resources \citep{ecb_2018}.} Activity logging and rapid access-revocation procedures can help detect and respond to unintended access \citep{nist_2020}. Developers could also monitor evaluators' activity against the approved scope of the review, and screen material before it leaves the company, as is done in trusted research environments for sensitive health and statistical data \citep{desai_2016, rds_nd, ndcal_nd}.\footnote{In patent litigation, source code is reviewed on the producing party's secured computers at counsel's offices, with no copies removed and every inspection logged \citep{ndcal_nd}.} Company-managed devices, non-disclosure agreements (NDAs), and counsel-reviewed or escrowed notes can also be used to restrict information egress \citep{dod_2012, ecb_2018}.\footnote{In US facilities accredited for sensitive compartmented information, visitors surrender personal electronic devices \citep{dod_2012}. European Central Bank inspection teams work in lockable secured offices, and external team members sign individual confidentiality agreements \citep{ecb_2018}.} To avoid unnecessary information being acquired by embedded evaluators incidentally, developers can ensure that evaluators are physically separated from operations that are not relevant to the assessment \citep{iba_2018, mckinsey_2005}.

Personnel vetting and employment cooling-off periods may further reduce the risk of sensitive knowledge being transferred to competitors (\citealp{iaea_1972}; \href{https://www.law.cornell.edu/uscode/text/12/1820}{12 U.S.C. § 1820(k)}; \href{https://www.law.cornell.edu/uscode/text/15/78j-1}{15 U.S.C. § 78j-1}).\footnote{States must consent to the designation of individual IAEA safeguards inspectors, in part to protect industrial secrets \citep{iaea_1972}. In auditing, senior audit staff cannot take senior financial roles at a client within one year (\href{https://www.law.cornell.edu/uscode/text/15/78j-1}{15 U.S.C. § 78j-1(l)}), and senior bank examiners face a one-year bar on accepting compensation from institutions they examined (\href{https://www.law.cornell.edu/uscode/text/12/1820}{12 U.S.C. § 1820(k)}).} In particular, cooling-off periods before evaluators are able to join one of the developers' competitors could reduce the risk of knowledge being transferred. Similar restrictions are used in financial supervision. For example, in the US, certain senior bank examiners face a one-year restriction on accepting paid employment or consulting work from institutions they have supervised (\href{https://www.law.cornell.edu/uscode/text/12/1820\#k}{12 U.S.C. § 1820(k)}). With sufficiently strong restrictions, embedded evaluators could in some respects pose less IP-transfer risk than employees of the developer, who may be able to move directly to competitors.

\textbf{Developers may be concerned that embedded assessments will identify improprieties that could pose reputational and legal risks.} Embedded evaluators may be more likely than other evaluators to identify findings that developers do not want published. For example, evaluators might discover that an internal monitoring system failed to detect unauthorized agent activity, despite the developer's public assurances of its effectiveness. Evaluators may also become holders of legally relevant information, which could trigger reporting obligations or make their records subject to disclosure in legal proceedings, depending on the applicable law and the evaluator's role. For example, information collected during audits can be subject to compelled disclosure, as was the case when a federal court ordered Capital One to produce the forensic breach report that a security consultant had prepared \citep{carroll_2020}. If embedded assessments remain voluntary, they could therefore subject well-intentioned actors to additional reputational or legal risks, while shielding developers that do not participate. This could have a chilling effect on embedded assessments, reducing the number of developers willing to engage with evaluators or the depth of access they are willing to allow.

To a large extent, these are risks that developers will have to incur. The purpose of embedded assessments is to provide assurance about the risks posed by a developer's activities, which necessarily creates the possibility that improprieties will be identified and disclosed. At the same time, developers may gain reputational benefits from subjecting themselves to credible independent scrutiny and publicly demonstrating how they respond to weaknesses that evaluators identify. There are also several ways for developers to manage reputational and legal risks. First, developers and evaluators can agree on procedures for handling and publishing findings \citep{longpre_2025}. For example, sensitive information could be redacted from public reports, as is common in government, nuclear, and pharmaceutical audits (\citealp{gao_2021}; \href{https://www.ecfr.gov/current/title-10/chapter-I/part-2/subpart-C/section-2.390}{10 C.F.R. § 2.390}; \href{https://www.ecfr.gov/current/title-21/chapter-I/subchapter-A/part-20/subpart-D/section-20.61}{21 C.F.R. § 20.61}), as well as in current AI evaluations \citep{metr_2026c}.\footnote{Redactions should protect sensitive information without concealing findings merely because they are unfavorable to developers. Information clearly outside the assessment's scope could also be omitted, provided this does not obscure material safety concerns or undermine the report's conclusions.} Reviews could also take place under NDAs that reduce the risk of information becoming public, as is required of financial auditors under professional conduct rules \citep[§1.700.001]{aicpa_nd} and is already used in AI evaluations. Developers could also be given an opportunity to review the findings for factual accuracy, publish a response to unresolved disagreements, and issue a follow-up report explaining how they intend to address any concerning findings. Exit clauses could even allow developers to withdraw from an evaluation before findings are published. Second, developers and evaluators can establish procedures for handling legally compelled disclosure. These could require evaluators to notify the developer of any disclosure request, unless legally prohibited from doing so, and to seek protective orders or confidential treatment for sensitive information where available.

\textbf{Evaluators and developers would need to dedicate time and resources to embedded assessments.} Developers would need to devote staff time to setting up and supporting an assessment. For example, evaluators may require vetting, company-managed devices, supervised workspaces, and escorted access to sensitive systems and facilities. Establishing and maintaining these arrangements can draw on engineering, security, legal, and management capacity throughout the engagement \citep{homewood_2025}. That said, many of these requirements overlap with those for routine employee onboarding, and we expect the costs to be substantially outweighed by the value of identifying important risks or vulnerabilities. Some costs to developers may also be offset where the assessment generates evidence that supports reporting or assurance obligations, such as internal-use risk reporting under SB 53 \citep{california_2025}.\footnote{Under SB 53 § 22757.12(a)(4), frontier developers are required to ``transmit to the Office of Emergency Services a summary of any assessment of catastrophic risk resulting from internal use of its frontier models every three months or pursuant to another reasonable schedule''.} The relative resource burden is likely to be greater for evaluators, given the relatively small number of qualified organizations able to conduct these reviews \citep{brundage_2026}. Evaluation organizations may therefore need to reallocate staff away from other activities, such as pre-deployment evaluations and research. Over time, however, the growing demand for embedded assessments may help expand evaluator capacity.

Several design choices could help keep these demands manageable. Standardized methods and shared resources, such as contractual templates and knowledge-sharing between evaluators, could reduce the effort required to set up each assessment. Developers can also draw on practices used in other industries, such as appointing a dedicated liaison to coordinate the assessment \citep{ismael_2018}. On the evaluator side, resource demands could also be reduced by limiting the duration of the assessment or the number of evaluators involved, though this would come at the cost of a less thorough review.

The challenges described above may raise concerns among developers considering embedded assessments. However, it seems possible to address these concerns through careful assessment design and appropriate safeguards. Overall, we think the benefits of embedded assessments substantially outweigh the risks.

\section{Key design choices for embedded assessments}\label{sec:3}

In this section, we explore seven questions about how embedded assessments could be designed. They cover the main choices that developers and evaluators would need to agree on before an evaluation begins. We identified them by drawing on previous evaluations for frontier AI and evaluations conducted in other high-stakes industries.

We ask what systems and practices evaluators could examine (\Cref{sec:3.1}), how evaluators could gather information (\Cref{sec:3.2}), how long assessments could last (\Cref{sec:3.3}), when they could take place (\Cref{sec:3.4}), what terms could govern the assessment (\Cref{sec:3.5}), how findings could be disclosed (\Cref{sec:3.6}), and what escalation procedures could be put in place (\Cref{sec:3.7}).

Note that \Cref{sec:3.1} focuses on a specific use case for embedded assessments -- risks from developers' internal AI use -- whereas the other sections consider design choices that apply to embedded assessments generally.

\subsection{What systems and practices could evaluators examine?}\label{sec:3.1}

Embedded assessments could be used to assess a range of risk areas. In this paper, we focus on their use in assessing the risks from internal AI use, and the adequacy and effectiveness of the safeguards intended to manage those risks. Internal AI use includes internal testing and the use of models to write code, run experiments, or train, evaluate, and monitor other models. We think this is a high-priority area because developers increasingly deploy their most capable models internally to assist with or lead on increasingly complex tasks \citep{anthropic_2026k}. These uses create distinct risks because models are often permitted to operate with high autonomy, access sensitive systems and information, and take actions that directly affect the developer's infrastructure or the development of future models \citep{chan_2025}. Assessments of internal AI use risks are also particularly tractable in the near term given they have been the focus of pilot embedded assessments. At the same time, they are difficult to meaningfully assess through existing forms of external oversight because doing so often requires direct access to developers' internal systems and practices.

We focus on three aspects of internal AI use: internal agent monitoring (\Cref{sec:3.1.1}), internal agent security controls and permissions (\Cref{sec:3.1.2}), and model alignment (\Cref{sec:3.1.3}).\footnote{There are other assessment targets for internal use risks that we do not cover here, such as models being used to evaluate other models and design research experiments.} Within each, we identify some possible assessment targets, but these are not meant to be exhaustive.

Embedded assessments may also be used to assess a range of other risk areas, many of which are beyond the scope of this paper, including a developer's security mitigations to protect sensitive assets (e.g. model weights) or compliance with potential future pacing commitments (see the Appendix).

\subsubsection{Internal agent monitoring}\label{sec:3.1.1}

Embedded assessments could assess whether harmful agent behavior would be detected, escalated, and reviewed quickly enough to prevent serious harm. This is particularly important because internal agents may operate with fewer guardrails than public deployments. They may also have access to safety-critical assets and infrastructure, including model weights, safety evaluation pipelines, production codebases, and the ability to launch training runs \citep{delaney_2026, kwon_2026}. As a result, there could be serious consequences if internal agents were to pursue misaligned goals, for example by sabotaging assigned work, evading oversight, or operating autonomously beyond their intended scope \citep{anthropic_2026b, chan_2025, metr_2026b, openai_2026d}. Evaluators could assess whether monitoring systems provide sufficient coverage, accurately identify harmful behavior, and trigger appropriate escalation and review. \Cref{tab:1} summarizes these assessment targets and gives examples of the artifacts evaluators could review for each.

\textbf{Monitor coverage.} Monitors can only catch and escalate misbehavior in the traffic they cover. Because monitoring all traffic with the most capable monitors is very costly, developers often apply cheaper monitors broadly and reserve stronger monitors for higher-risk activities \citep{anthropic_2024, cunningham_2025, openai_2026e}. Coverage gaps can therefore arise in several ways. For example, some traffic may be exempt from monitoring entirely, triage criteria may inadvertently route high-risk activity to weaker monitors, or monitoring records may be incomplete.

There is evidence that such gaps arise in practice. For example, METR's pilot embedded assessment found that Anthropic's internal monitoring did not cover non-Claude Code traffic and missed subagent calls entirely \citep{metr_2026c}. Developers do not currently disclose what fraction of internal agent activity is monitored, meaning some agents may be deployed without appropriate monitoring \citep{jorgensen_2026, lindner_2025}. Further, high reported coverage within one category of activity can still leave important gaps elsewhere. For example, OpenAI reported that only a small fraction of traffic (\textless{}0.1\%) is not covered by monitoring \citep{openai_2026h}. However, the Hugging Face incident caused by internal OpenAI models was not identified due to a gap in the monitoring set-up -- OpenAI's chain-of-thought monitors did not run on the cybersecurity evaluations involved \citep{openai_2026b}.\footnote{OpenAI subsequently expanded its monitoring requirements to cover tool-using training and evaluations involving sufficiently capable models \citep{openai_2026e}.} Incomplete records can create further gaps in scrutiny. In METR and Redwood's investigation of the same incident, around 10\% of agent activity was estimated to be missing from transcripts due to container resets and early terminations \citep{metr_2026d}.

Embedded assessments could help determine whether such gaps exist. Evaluators could assess the extent to which consequential internal agent traffic -- such as subagent calls, evaluation runs, and agent activity involving sensitive data, production systems, model weights, training infrastructure, or other safety-critical resources -- is subject to monitoring. As part of this, they could test whether agents can cause their own traffic to bypass monitoring. They could also assess whether the criteria used to triage traffic between stronger and weaker monitors capture the highest-risk activity, such as agents with access to safety-critical systems.

\textbf{Monitor accuracy.} Even where traffic is covered, monitors may not reliably identify every safety-relevant action \citep{guan_2025}, especially where agents actively try to conceal or evade detection \citep{lindner_2025}. Effective monitoring should minimize both false negatives (where harmful behavior goes undetected) and false positives (where benign behavior is incorrectly flagged).\footnote{Monitoring may also become substantially less useful if models can control their chain-of-thought, or if future models use a ``neuralese'' to reason in forms not legible to humans.}

Recent evidence suggests these failures can occur in practice. METR's pilot embedded assessment found that basic jailbreaks reliably fooled Anthropic's monitors, even when the monitor was warned about the attack \citep{metr_2026c}. Monitors used during model training show similar limitations. Anthropic red-teamed its reinforcement learning monitoring on known reward hacks and found that around 90\% were reliably detected.\footnote{Anthropic also stress-tested this monitoring by prompting the model to iteratively rewrite detected hacks while looking at the monitor's reasoning between rounds. Around 40\% of tested hacks fell below the severity threshold, and Anthropic concluded that its monitoring was robust to one-shot obfuscation attempts but not adversarial iteration \citep{anthropic_2026c}. Anthropic notes that it does not train against this monitoring and considers the iterative-rewriting condition unrealistically adversarial.} One gap involved a hack in which the model stated a wrong answer confidently enough that a grader model accepted it \citep{anthropic_2026c}. Monitors can also be fed tampered records. During the OpenAI-Hugging Face incident, agents learned to spoof their tool calls, so the commands and outputs recorded in transcripts no longer matched their actions. METR and Redwood Research found that tool-call spoofing techniques affected roughly 7\% of the transcripts they reviewed \citep{metr_2026d}.

Evaluators could assess monitor accuracy in several ways. They could verify the accuracy of the developer's own reported monitor-accuracy scores.\footnote{Anthropic has reported catch rates above 80\% \citep{anthropic_2026c} and OpenAI has reported that its monitors outperform employee escalations \citep{openai_2026h}.} They could also test whether monitors reliably flag harmful agent behavior, including behavior that agents attempt to conceal. Evaluators could also measure how architecture changes affect monitorability -- for example, whether a new architecture gives models more control over their chain-of-thought \citep{redwood_2026}. In addition, evaluators could assess whether internal agent monitors have access to an agent's reasoning, and the extent to which this improves detection.\footnote{For instance, METR found that reasoning access increased average catch rates from 52\% to 80\% \citep{kinniment_2026}.}

Recent empirical evidence shows that there are also broader concerns about the future monitorability of model reasoning. OpenAI's Astra is substantially better than GPT-5.6 at completing tasks without using chain-of-thought, and is also better at controlling its chain-of-thought \citep{openaiaisi_2026}. Additionally, OpenAI found that ``when prompted Astra is able to strategically underperform evaluations (sandbagging) in ways that evade even sandbagging-specific monitors'' \citep{openai_2026f}. Anthropic has separately reported that a technical error in how its reward signal was configured during reinforcement learning meant that rewards were partly based on scratchpad content in 8\% of training episodes. This put training pressure on the model's reasoning traces that could undermine the monitorability of its chain-of-thought \citep{anthropic_2026c}. Evaluators could therefore test whether chain-of-thought remains a reliable signal of harmful behavior, including when agents attempt to evade monitoring, and examine whether training practices undermine that reliability \citep{transluce_2026}.

\textbf{Escalation and review.} Detecting harmful behavior is only useful if the resulting flag reaches an appropriate reviewer, whether automated or human. This can fail if later filters discard trajectories that were initially flagged, meaning harmful behavior is detected but never reviewed \citep{jorgensen_2026}. Review also needs to happen quickly enough to prevent harm. Because monitoring is often offline or asynchronous, harmful actions may already have occurred by the time a trajectory is reviewed \citep{jorgensen_2026, lindner_2025}.

It is unclear whether developers currently evaluate escalation and review practices thoroughly enough. Anthropic notes that it has not extensively evaluated the summarization and human-review stages of its monitoring pipeline, and that it occasionally updates filters without evaluating the effect on recall \citep{anthropic_2026e}. Unreviewed flags may be particularly consequential during training because harmful or anomalous behavior can propagate into model weights, datasets, or later stages of the training pipeline. Developers use models to screen training environments and monitor training episodes, but human review covers only part of what these systems flag \citep{anthropic_2026c}.\footnote{Anthropic screens training environments before use, including a dedicated evaluation for power-seeking incentives in which models score environments and humans manually review the highest-scoring ones \citep{anthropic_2026c}. It also uses visibility tools that surface potential prompt-reward mismatches for researcher attention \citep{anthropic_2026f}. However, Anthropic notes that it has not quantified the effectiveness of its environment evaluation pipeline and that the process did not catch every problematic environment in time to avoid training on it.}

Embedded evaluators could therefore assess whether flagged trajectories are escalated to appropriate reviewers and reviewed quickly enough to prevent harm. For training specifically, evaluators could assess whether training-time flags reach human reviewers in time to affect the run, and whether the flagged behavior was harmful (or could have led to harm), for example by re-examining flagged environments and episodes.

\renewcommand{\arraystretch}{1.2}
\begin{xltabular}{\linewidth}{%
        >{\small\raggedright\arraybackslash}p{\dimexpr0.2333\linewidth-2\tabcolsep\relax}%
        >{\small\raggedright\arraybackslash}p{\dimexpr0.3333\linewidth-2\tabcolsep\relax}%
        >{\small\raggedright\arraybackslash}X}
    \toprule
        \textbf{Assessment target} &
        \textbf{Description} &
        \textbf{Example artifacts}\\
    \midrule
    \endfirsthead
    \toprule
        \textbf{Assessment target} &
        \textbf{Description} &
        \textbf{Example artifacts}\\
    \midrule
    \endhead
    \bottomrule
    \endfoot
    \bottomrule
    \addlinespace[0.5em]
    \caption{Assessment targets and artifacts for internal agent monitoring}
    \label{tab:1}\\
    \endlastfoot
        \textbf{Monitor coverage} &
        Evaluators could assess the extent to which consequential agent traffic (e.g. subagent calls, evaluation runs, and agent activity involving sensitive data, production systems, model weights, training infrastructure, or other safety-critical resources) is monitored &
        \vspace{-0.7em}\begin{itemize}[leftmargin=*, nosep]
\item Live monitoring pipeline (i.e. running agent traffic through the monitoring pipeline to see if it is monitored)
\item Monitor coverage figures (i.e. what fraction of internal agent traffic is monitored versus exempted)
\item Monitor assignment criteria (i.e. the rules determining which traffic is routed to stronger or weaker monitors)
\item Traffic attribution logic (i.e. the logic used to determine whether an action gets monitored)
\item Sample agent transcripts (i.e. which parts of agent activity were covered by monitoring)
\end{itemize}\vspace{-0.2em}\\
        \textbf{Monitor accuracy} &
        Evaluators could assess whether monitors reliably flag harmful agent behavior, including behavior agents attempt to conceal &
        \vspace{-0.7em}\begin{itemize}[leftmargin=*, nosep]
\item Live monitors (i.e. red-teaming monitors while they run to test how easily they can be jailbroken)
\item Classifiers (i.e. reviewing how the classifiers screening internal agent traffic are configured, and testing whether that configuration catches misuse)
\item Chain-of-thought monitors (i.e. testing how reliably monitors that read model reasoning catch deception)
\item Sample flagged trajectories (i.e. reviewing how often flags prove to be false positives)
\item Monitoring effectiveness assessments (i.e. checking whether reported catch rates and false positive rates are accurate)
\end{itemize}\vspace{-0.2em}\\
        \textbf{Escalation and review} &
        Evaluators could assess whether flagged trajectories are escalated to appropriate reviewers and reviewed quickly enough to prevent harm &
        \vspace{-0.7em}\begin{itemize}[leftmargin=*, nosep]
\item Escalation logic and queues (i.e. reviewing which flags are escalated to reviewers, and how quickly they are reviewed)
\item Review pipeline filters (i.e. checking how many initially flagged trajectories are discarded before reaching a reviewer)
\item Incident logs (i.e. reviewing how past flags were handled and whether review occurred quickly enough to prevent harm)
\end{itemize}\vspace{-0.2em}\\
\end{xltabular}

METR's pilot suggests that internal agent monitoring is a valuable and highly tractable focus for embedded assessments. The pilot covered most of the assessment targets we set out for monitor coverage and monitor accuracy, showing that these questions can be investigated in practice through embedded access \citep{metr_2026c}.\footnote{See METR's Frontier Risk Report \citep{metr_2026b} (Appendix B, pp. 74--76). The red-team exercise found it ``relatively simple to change an environment variable to register the traffic as not coming from Claude Code'' (traffic attribution); identified a bug meaning ``subagent calls weren't being monitored'' (coverage); and found basic monitor jailbreaks that were ``extremely effective and robust'' (robustness). However, the report's account of the exercise did not mention an assessment of escalation and review.} Extending future assessments to escalation and review could provide additional value by testing not only whether harmful behavior is detected, but whether resulting flags are acted on appropriately.

However, assessments of internal agent monitoring also have important limitations. Findings may become outdated relatively quickly because monitoring set-ups change continuously alongside internal deployments \citep{kwon_2026}. As a result, periodic assessments may be insufficient. METR's pilot lasted three weeks while covering only a subset of the assessment targets we propose here \citep{metr_2026c}. Assessments that also examine escalation and review, or that aim to verify developers' own monitoring records and logs, may take even longer and impose significant compute costs. For example, METR and Redwood Research's review of agent behavior through roughly 1,300 transcripts and 1.2 million message board entries in the OpenAI-Hugging Face incident grew from two planned onsite days to six, and their AI-assisted analysis cost roughly \$400,000 in API credits \citep{metr_2026d}.

\subsubsection{Internal agent security controls and permissions}\label{sec:3.1.2}

Embedded assessments could examine whether internal agents are appropriately constrained even when monitoring fails. Security controls can contain agents within sandboxes, restrict the tools and websites they can access, block access to sensitive assets (e.g. model weights), and deny the compute or credentials needed to operate autonomously without human knowledge or permission \citep{metr_2026b}. Recent incidents have shown that agents can escape sandboxes and exploit access to sensitive infrastructure where sufficient controls are not in place \citep{npr_2026, openai_2026d}. Evaluators could assess whether a developer's controls are effective by evaluating agent permissions and security controls. \Cref{tab:2} summarizes these assessment targets and gives examples of the artifacts evaluators could review for each.

\textbf{Security controls.} Security controls such as sandboxing, filesystem restrictions, and network egress policies determine what an agent can reach and whether data can leave its environment, even if an agent acts beyond its intended permissions \citep{fmf_2026}. If these controls are weak, inconsistently applied, or misconfigured, they cannot serve as an effective backstop when monitoring fails.

Recent incidents show that these controls can fail in practice. Internal OpenAI models escaped their evaluation sandbox through a vulnerable package proxy, while a misconfigured sandbox allowed one of Meta's internal models to reach the internet \citep{npr_2026, openai_2026d}. Similarly, evaluation machines in Anthropic's cybersecurity testing were misconfigured with live internet access, allowing models intended to be isolated to compromise the production systems of three organizations \citep{anthropic_2026b}. Current practice also varies in how consistently internal agent controls are applied. Anthropic reports that its internal agents run with combinations of sandboxing and website whitelists, but that only some agents are sandboxed \citep{anthropic_2026f}. OpenAI requires isolation for workloads that execute model-generated or otherwise untrusted code \citep{openai_2026e}.

Evaluators could assess how these controls are applied in practice. They could review the basis for differences in how internal agents are constrained, assess whether developers' policies are consistently followed and executed, and examine whether unsandboxed agents perform tasks or have access that warrants stronger containment. Evaluators could also test sandboxes directly by instructing capable models to attempt escapes from within production sandboxes, and by red-teaming isolation controls to determine whether an agent that compromises its workload or a supporting service -- such as a package proxy or evaluation harness -- can gain unauthorized access to the internet or other internal networks. They could also assess whether attempted and successful escapes are detected, including how long successful escapes remain undetected.

\textbf{Agent affordances, permissions, and credentials.} Agents should receive only the affordances (i.e. the tools and websites they can use) and permissions (i.e. the actions they are authorized to take) needed to complete their assigned tasks. This can reduce the risk that agents take actions beyond their intended scope \citep{apollo_2025, google_2025, pistillo_2026}. Agents' credentials (i.e. the API keys and authentication tokens they use to access other systems) should be similarly scoped, as should their access to the data and compute that their assigned tasks require \citep{fmf_2026}. Unauthorized access to compute is particularly important because it could allow an agent to sustain or scale its activity and operate autonomously without human knowledge or permission.\footnote{METR noted from its cross-developer assessment that developers vary in how thoroughly they track internal compute usage \citep{metr_2026b}.}

Developers have begun developing practices to enforce these limits. One approach is to restrict tool access on a per-task basis \citep{fmf_2026}. Another is to govern individual actions at runtime. For example, Google describes policy engines that evaluate each proposed action against predefined rules and then allow it, block it, or require user confirmation \citep{google_2025}.

Embedded evaluators could assess whether these limits are appropriately designed and enforced. They could determine whether agents' affordances and permissions are deliberately scoped to their tasks, and whether agents can obtain credentials, compute, or data beyond what their tasks require. Evaluators could also review how permission increases are granted to reduce the risk that agents are given affordances, permissions, or credentials beyond what is necessary for their tasks \citep{apollo_2025}.

\renewcommand{\arraystretch}{1.2}
\begin{xltabular}{\linewidth}{%
        >{\small\raggedright\arraybackslash}p{\dimexpr0.2333\linewidth-2\tabcolsep\relax}%
        >{\small\raggedright\arraybackslash}p{\dimexpr0.3333\linewidth-2\tabcolsep\relax}%
        >{\small\raggedright\arraybackslash}X}
    \toprule
        \textbf{Assessment target} &
        \textbf{Description} &
        \textbf{Example artifacts}\\
    \midrule
    \endfirsthead
    \toprule
        \textbf{Assessment target} &
        \textbf{Description} &
        \textbf{Example artifacts}\\
    \midrule
    \endhead
    \bottomrule
    \endfoot
    \bottomrule
    \addlinespace[0.5em]
    \caption{Assessment targets and artifacts for internal agent security controls and permissions}
    \label{tab:2}\\
    \endlastfoot
        \textbf{Security controls} &
        Evaluators could assess whether sandboxes and other isolation measures effectively contain agents &
        \vspace{-0.7em}\begin{itemize}[leftmargin=*, nosep]
\item Agent sandboxes (i.e. testing live how easily sandboxes can be escaped)
\item Sandboxing criteria (i.e. reviewing how developers decide which agents and workloads require sandboxing, and whether these criteria are followed in practice)
\item Egress restrictions (i.e. reviewing what can leave an agent's environment, and whether escapes would be detected)
\item Isolation controls (i.e. testing whether an agent that compromises a workload can reach the internet or other internal networks)
\item Agent activity logs (i.e. reviewing whether agents' actions and planning remain observable)
\end{itemize}\vspace{-0.2em}\\
        \textbf{Agent affordances, permissions, and credentials} &
        Evaluators could assess whether agents' affordances, permissions, and credentials are appropriately scoped and reviewed, and whether agents can access tools, compute, or data beyond what their tasks require &
        \vspace{-0.7em}\begin{itemize}[leftmargin=*, nosep]
\item Compute credential provisioning (i.e. testing whether agents could obtain compute they were not granted, and how usage is tracked)
\item Internal system access records (i.e. reviewing what infrastructure and data agents can actually reach)
\item Agent permission configurations (i.e. reviewing what tools and access internal agents have, including website whitelists)
\item Runtime policy engines (i.e. reviewing the rules used to allow, block, or require confirmation for agent actions)
\item Permission change records (i.e. reviewing how permission increases are requested, approved, and reviewed)
\end{itemize}\vspace{-0.2em}\\
\end{xltabular}

Recent incidents have illustrated the need for independent scrutiny of internal agent security controls and permissions. Following the OpenAI-Hugging Face incident, in which models escaped their sandbox and compromised Hugging Face's production infrastructure, OpenAI agreed to an embedded assessment in the form of a post-incident review by METR and Redwood Research \citep{metr_2026g, openai_2026d}. Future embedded assessments could also provide scrutiny before incidents occur, rather than after (see \Cref{sec:2.1}).

However, assessments of security controls and permissions may require access to particularly sensitive infrastructure. Sandbox configurations, access controls, and egress policies present information security risks, so developers may be reluctant to provide third parties with access, even if they are onsite (see \Cref{sec:2.2}). It is also unclear how such assessments would work in practice, as they have not yet been trialed, to our knowledge. That said, it seems possible to mitigate these challenges by implementing appropriate safeguards.

\subsubsection{Model alignment}\label{sec:3.1.3}

Embedded assessments could examine whether models remain aligned throughout their lifecycle and whether current alignment techniques are likely to remain effective as models become more capable. This is particularly important as frontier AI developers increasingly use their most capable models to help develop and align future systems \citep{anthropic_2026g}. If alignment techniques fail to keep pace with model capabilities, misalignment could emerge during model development and deployment.

Evaluators could use two complementary approaches. First, they could evaluate models across their lifecycle: training-run evaluations could search for concerning behavior in training rollouts, pre-deployment evaluations could assess the model's final checkpoint before release, and deployment monitoring could examine models as they are used internally or externally. We focus on training-run evaluations and deployment monitoring, since these are the stages where embedded access is most relevant.\footnote{Pre-deployment evaluations remain important but typically require only inference API access to the final checkpoint, rather than full embedded access, and are therefore outside the scope of this paper.} Second, evaluators could assess whether current alignment techniques and processes are likely to remain effective on more capable near-future models. \Cref{tab:3} summarizes these assessment targets and gives examples of the artifacts evaluators could review for each.

\textbf{Training incentives and emerging misalignment.} Training environments can incentivize undesirable behavior when their reward functions are misspecified or underspecified. For example, models may inadvertently learn to engage in power-seeking behavior, which involves taking actions to acquire resources or influence as an instrumental or intermediary goal toward achieving a longer-term goal \citep{anthropic_2026c}. Training configuration errors can also undermine safety measures, for example by placing reward pressure on model reasoning, therefore incentivizing models to be less transparent in their reasoning \citep{anthropic_2026c, carroll_2026}. Scheming-related behavior, where a model ostensibly follows a user's instructions while covertly pursuing a secondary, misaligned goal, can also increase during training and then stop being visible if the model internalizes its reasoning or learns to state it innocuously \citep{apollo_2026}. Evaluations of the final model alone could miss this, because the released checkpoint may show no visible trace of behavior that was detectable mid-training \citep{apollo_2026}.

There is evidence that these risks can arise in practice. Models that learned to reward hack production coding environments have generalized to behaviors such as alignment faking and sabotage \citep{denison_2024, macdiarmid_2025}. Anthropic's Claude 3 Opus exhibited alignment faking in training experiments, strategically complying with its training objective to avoid modification of its behavior \citep{greenblatt_2024}. Moreover, models have been observed reasoning about their graders and reward processes at checkpoints across a production reinforcement learning run. For example, verbalized evaluation awareness rose from 2\% to over 20\% during capabilities-focused reinforcement learning (RL) training of OpenAI's o3 \citep{schoen_2026}. Developers' own review processes may not reliably catch all such problems. For example, while Anthropic conducts pre-training evaluations, including dedicated reviews for environments that might incentivize power-seeking, it notes that it has not quantified the effectiveness of the pipeline and that some problematic environments were not caught before training \citep{anthropic_2026c}.

Hardening environments against exploits has also been shown to substantially reduce reward hacking \citep{thaman_2026}. For example, an Anthropic employee interviewed by METR described strengthening environments as the ``[m]ain way we've defended against reward hacks'' \citep{metr_2026b}. Embedded evaluators could assess a developer's practices for testing whether training environments incentivize harmful behavior, and whether training configurations undermine safety measures. For example, they could test whether models are rewarded for exploiting flaws in grading or environments rather than completing the task as specified \citep{anthropic_2026c, metr_2026b, transluce_2026}. Evaluators could also assess how effectively developers harden these environments against exploits and the extent to which these techniques reduce misbehavior. To do so, evaluators could use techniques like synthetic document fine-tuning to measure reward-seeking in intermediate checkpoints \citep{hojmark_2026} and insert honeypot beliefs to test model alignment \citep{wang_2025}. Finally, evaluators could examine intervening training data and environments to better understand what caused increases in scheming-related behavior \citep{apollo_2026}.

\textbf{Misaligned behavior during deployment.} Embedded assessments could continue during deployment because some concerning behavior may only emerge when models encounter out-of-distribution environments not seen during training.

Monitoring deployed models can reveal patterns of behavior that may not be apparent in pre-deployment evaluations. For example, model-organism research has shown that models can sandbag by hiding dangerous capabilities during evaluations \citep{vanderweij_2024}. Safety techniques, such as inoculation prompting, can also break down outside the conditions in which they were tested \citep{dubinski_2026}, while models may influence users in systematic ways that are difficult to detect from individual interactions or chain-of-thought alone \citep{betley_2026}.

Evaluators could take a similar approach to that described for internal agent monitoring in \Cref{sec:3.1.1}. For example, diffuse control monitoring \citep{hebbar_2025} involves running model graders across large quantities of deployment transcripts to calculate aggregate metrics, while requiring human inspection of a smaller sample of transcripts.

\textbf{Generalization of alignment techniques to near-future models.} Embedded evaluators could also assess whether current alignment techniques are likely to remain effective as models become more capable. Developers expect to outsource an increasing amount of capabilities and alignment work to automated AI researchers \citep{anthropic_2026g, openai_2026g}. However, AI systems trained using reinforcement learning with verifiable rewards may become better at capabilities research than alignment research, creating a risk that an automated alignment researcher may fail to keep pace with advances in AI capabilities \citep{bowkis_2026}. Misalignment in an automated researcher could also propagate to future models, particularly as fewer humans are in the loop to check their work \citep{hallensleben_2026}. Developers also make extensive use of their own models as graders and scorers during training \citep{anthropic_2026c, metr_2026b}. These models can be fooled or systematically biased, and extensive use of models as scorers increases the risk that scorer models cooperate with the model being trained to conceal misalignment \citep{anthropic_2026c}.

Several empirical observations also motivate closer scrutiny of alignment techniques. OpenAI's o3 model (released in 2025) demonstrated increased reward seeking during capabilities training \citep{hojmark_2026}. This tendency seems not to have been fully fixed in later models, since the same drive for ``understanding the implementation of the scorer'' was identified as one of the main motivations for the models in the OpenAI-Hugging Face incident \citep{metr_2026d}. Previous work has also shown that once a model develops misaligned behavior, this can persist through standard alignment training \citep{hubinger_2024}. Anthropic's ``Hacker-Opus'' reward-seeking model provides another example \citep{anthropic_2026h}. Anthropic's automated alignment evaluations did not detect severe misalignment, even though the checkpoint showed a strong propensity to participate in a simulated Hugging Face attack \citep{hubinger_2026}.

Evaluators could assess these risks in several ways. They could stress-test new alignment techniques using affordances comparable to those available to internal alignment researchers, providing an independent second opinion and identifying potential weaknesses. They could also examine procedural assumptions that may break down for more capable near-future models (e.g. whether it remains safe to beta-test a possibly misaligned model internally). Finally, they could interview technical staff about whether they believe current alignment techniques will remain reliable over the next three, six, and twelve months and -- where gaps are anticipated -- whether replacement techniques are likely to be ready in those time periods.

\renewcommand{\arraystretch}{1.2}
\begin{xltabular}{\linewidth}{%
        >{\small\raggedright\arraybackslash}p{\dimexpr0.2333\linewidth-2\tabcolsep\relax}%
        >{\small\raggedright\arraybackslash}p{\dimexpr0.3333\linewidth-2\tabcolsep\relax}%
        >{\small\raggedright\arraybackslash}X}
    \toprule
        \textbf{Assessment target} &
        \textbf{Description} &
        \textbf{Example artifacts}\\
    \midrule
    \endfirsthead
    \toprule
        \textbf{Assessment target} &
        \textbf{Description} &
        \textbf{Example artifacts}\\
    \midrule
    \endhead
    \bottomrule
    \endfoot
    \bottomrule
    \addlinespace[0.5em]
    \caption{Assessment targets and artifacts for model alignment}
    \label{tab:3}\\
    \endlastfoot
        \textbf{Training incentives and emerging misalignment} &
        Evaluators could assess intermediate checkpoints for concerning propensities, such as scheming, reward-seeking, and evaluation awareness, and track how these change over the course of training. They could also measure whether train-time monitors detect anti-corrigible behavior like alignment faking. &
        \vspace{-0.7em}\begin{itemize}[leftmargin=*, nosep]
\item Intermediate model checkpoints (i.e. evaluating various model versions saved during a training run to search for concerning propensities that may not be observed in the final model checkpoint)
\item Developers' evaluation records across training checkpoints (i.e. reviewing the developer's own training-time evaluation results to identify concerning patterns, such as concerning propensities rising and then dramatically falling)
\item Checkpoint reasoning traces (i.e. assessing whether the model's chain of thought remains legible and continues to reflect the reasoning behind its behavior)
\item Randomly sampled RL environments (i.e. testing a random sample of environments directly to see whether they can be gamed or reward unintended behavior)
\item Access to a subset of RL environments, including for red-team/blue-team games to stress-test train-time monitors
\end{itemize}\vspace{-0.2em}\\
        \textbf{Misaligned behavior during deployment} &
        Evaluators could monitor internal and external traffic for signs of misalignment &
        \vspace{-0.7em}\begin{itemize}[leftmargin=*, nosep]
\item Sample agent transcripts (i.e. which parts of agent activity were covered by monitoring)
\item Classifiers (i.e. reviewing how the classifiers screening internal agent traffic are configured, and testing whether that configuration catches misuse)
\item Chain-of-thought monitors (i.e. testing how reliably monitors that read model reasoning catch deception)
\item Sample flagged trajectories (i.e. reviewing how often flags prove to be false positives)
\item Incident logs (i.e. reviewing how past flags were handled and whether review occurred quickly enough to prevent harm)
\end{itemize}\vspace{-0.2em}\\
        \textbf{Generalization of alignment techniques to near-future models} &
        Evaluators could determine whether current alignment techniques and procedures are expected to apply to models in three, six, and twelve months &
        \vspace{-0.7em}\begin{itemize}[leftmargin=*, nosep]
\item Internal procedures for mitigating misalignment risk (i.e. examining whether these procedures still hold up as capabilities increase, for example assessing whether internally beta-testing a possibly misaligned model remains safe, whether automated alignment evaluations would catch severe misalignment, or whether model graders can be trusted not to collude with the model being trained)
\item New alignment techniques and their supporting evaluations (i.e. independently stress-testing techniques before they are relied upon, providing a second opinion and identifying weaknesses)
\item Records on how quickly models are improving at capabilities research and alignment research (i.e. to assess whether alignment progress is keeping pace, including for automated researchers)
\item Transcripts of structured interviews with alignment researchers (i.e. eliciting predictions on whether current techniques will remain reliable over the next three, six, and twelve months and, where gaps are anticipated, whether replacements will be ready in time)
\end{itemize}\vspace{-0.2em}\\
\end{xltabular}

Embedded assessments could help to provide scrutiny of models intended solely for internal use, which may otherwise fall outside third-party evaluation processes triggered by planned public deployment. In particular, training-run evaluations could identify concerning behaviors that are not easily observable in the final model checkpoint. For example, reviewing a sample of training environments could surface environments that encourage concerning model propensities. Sampling may also impose relatively limited costs on developers because training data of this kind, while intellectually sensitive, is mainly of concern when shared in bulk \citep{apollo_2026}. Moreover, onsite access with company-managed devices further reduces the risk of data leaving the company.\footnote{Checkpoint access is also relatively inexpensive to provide, since developers can expose checkpoints to evaluators over an API in the same way as deployed models, and evaluations can run in parallel with training rather than delaying releases \citep{apollo_2026}.} Evaluating how well alignment techniques hold could provide another benefit: identifying weaknesses before they affect future frontier-scale training runs. Better information about how long current alignment techniques are likely to remain effective could also inform decisions about the pace of frontier development \citep{amodei_2026}.

However, these assessments would require access to particularly sensitive systems and substantial technical expertise. Thorough training-run evaluations could require access to commercially sensitive systems and methods, including training pipelines and reinforcement learning environments. At the same time, training-run evaluations need not initially impose large costs on developers: they can run in parallel with training rather than sit on the critical path to release, and early pilots could begin with evaluator-supplied tests that developers run on their own checkpoints before moving toward more direct inspection of training processes \citep{apollo_2026}. Risks associated with access to sensitive intellectual property could also be mitigated through measures such as sampling and the physical security controls available through onsite access. Assessing whether alignment techniques will generalize to more capable models may be even more demanding, requiring both substantial experimental resources and expertise in alignment techniques comparable to that of senior alignment researchers. Embedded access may nevertheless make this type of assessment more feasible than alternatives that require providing offsite evaluators with similarly extensive access to sensitive systems and methods.

\subsection{How could evaluators gather information?}\label{sec:3.2}

Embedded assessments can gather evidence through three main channels: direct access to infrastructure, engagement with personnel, and review of documentation (see \Cref{tab:4}). These sources provide complementary evidence and allow evaluators to cross-check findings across multiple sources. Most assessments will therefore need to combine all three.

\textbf{Infrastructure access.} Infrastructure access involves developers providing evaluators with controlled access to the internal systems used to deploy, monitor, and constrain internal AI models (e.g. monitoring pipelines, codebases, classifiers, and sandboxes). Evaluators would have access with parity to that of a member of the developer's internal risk-assessment team or the team responsible for the system being tested (e.g. a senior model alignment researcher). Evaluators should be able to use this access to directly test and red-team developers' safety systems.

The main benefit of infrastructure access is that it allows evaluators to test systems directly and assess how safeguards work in practice. This can reveal vulnerabilities that may not be captured by information reported by the developer. For example, in METR's pilot embedded assessment, METR noted that some vulnerabilities would have been ``highly non-obvious'' without specific technical knowledge of Anthropic's infrastructure \citep{metr_2026b}. Direct access can also allow evaluators to assess safeguards end to end. Rather than reviewing isolated artifacts, they can test monitor coverage, accuracy, and escalation together to assess whether the overall pipeline would detect and stop harmful behavior. Embedded evaluators can also test how components interact, which is important as some vulnerabilities may arise at the boundaries between systems. For example, in METR's pilot embedded assessment, METR observed a flaw where monitoring only covered traffic that the attribution logic labeled as coming from Claude Code, so changing a single environment variable relabeled agent traffic and routed it around monitoring entirely \citep{metr_2026b}.

However, infrastructure access can also create intellectual property, operational, and resource costs. It can expose proprietary systems and methods, including reinforcement learning environments, agent sandboxes, and affordance and permission controls, creating risks of model misuse or unintended transfer of highly valuable intellectual property. Live testing can also create operational risks. For example, red-teaming production monitors or sandboxes could trigger false alerts or disrupt real workflows. Containment measures could reduce this risk but would likely add coordination costs. For example, developers could tag evaluator traffic so that alerts can be filtered. Developers' systems may also take time to understand, and provisioning and supervising access can draw on engineering and security capacity.

\textbf{Personnel access.} Personnel access involves developers making relevant staff available for questioning and observation, for example through interviews, questionnaires, demonstrations, meetings, and internal messaging channels. Evaluators can use this access to question staff directly and observe how systems and processes work in practice.

A key benefit of personnel access is that it gives evaluators access to knowledge and context that is unlikely to be captured in formal documentation. Staff who build and operate safety systems may hold tacit and procedural knowledge that may not be formally recorded. Evaluator-designed interviews and questionnaires can capture practices that are not documented and reveal where staff see gaps in safety systems.\footnote{METR piloted this approach in its Frontier Risk Report, in which Anthropic, Google, Meta, and OpenAI responded to a detailed evaluator-designed questionnaire on model capabilities, internal AI use, and monitoring, surfacing information not available in public documentation \citep{metr_2026b}.} Direct staff access can also increase evaluators' confidence that important information is not being withheld, by allowing them to compare formal accounts with staff members' experiences and perspectives. It also gives evaluators flexibility to ask follow-up questions and pursue unexpected findings as they emerge. Furthermore, developers can give evaluators step-by-step demonstrations to show how systems work in practice and help evaluators design better tests. Evaluators can also benefit from attending relevant meetings as they can reveal how deployment, escalation, and incident decisions are made and whether these differ from written processes.

However, personnel access can also impose costs and create information-handling challenges. Interviews, questionnaires, and demonstrations take some staff time. Developers also have less control over what is disclosed when personnel access is provided. For example, staff may inadvertently reveal sensitive or out-of-scope information, including concerns they have not raised with management. There is also a risk that evaluator presence changes staff behavior, for example if employees self-censor or move sensitive discussions to channels evaluators cannot observe.\footnote{For example, employees may withhold or alter information they provide to reviewers due to psychological concerns, organizational influences, and legal considerations \citep{homewood_2025}.} Some forms of personnel access can also be difficult to scope. For example, internal messaging channels may combine safety-relevant material with unrelated sensitive content, personal information, or proprietary strategy that is not relevant to the assessment.

\textbf{Documentation review.} Documentation review involves sharing internal records used to support safety assessments and claims, such as monitoring-effectiveness assessments, incident logs, agent transcripts, safety cases, and internal policies. Evaluators can review the claims and methods in these documents and compare them with other evidence gathered during the evaluation.

The main benefit of documentation review is that it allows evaluators to scrutinize developers' own assessments and claims. For example, monitoring assessments can be used to check whether reported coverage and catch rates are supported by evidence. Documents can reveal gaps or methodological weaknesses in developers' own assessments. For example, monitoring-effectiveness assessments and evaluation transcripts may show that tests relied on weak attacks, missed parts of the pipeline, or otherwise overstated robustness. Documents can also help guide the rest of the evaluation. For example, agent logs and incident reports can show how systems and safeguards performed against real instances of problematic or unexpected behavior over time, including whether it was detected, escalated, and addressed. This can help evaluators identify weaknesses and decide which systems to red-team and which staff to interview.

However, documentation review can also create legal and resource costs. Sharing documents may create legal risks because board minutes, incident reports, and unredacted reports may become subject to compelled disclosure once shared \citep{carroll_2020}, which may make developers reluctant to provide them. Preparing documents for review can also be resource-intensive, as it may require legal and other staff to identify, redact, and review material for personal data and out-of-scope information, potentially slowing the evaluation.

\renewcommand{\arraystretch}{1.2}
\begin{xltabular}{\linewidth}{%
        >{\small\raggedright\arraybackslash}p{\dimexpr0.2333\linewidth-2\tabcolsep\relax}%
        >{\small\raggedright\arraybackslash}p{\dimexpr0.3333\linewidth-2\tabcolsep\relax}%
        >{\small\raggedright\arraybackslash}X}
    \toprule
        \textbf{Category} &
        \textbf{Description} &
        \textbf{Examples}\\
    \midrule
    \endfirsthead
    \toprule
        \textbf{Category} &
        \textbf{Description} &
        \textbf{Examples}\\
    \midrule
    \endhead
    \bottomrule
    \endfoot
    \bottomrule
    \addlinespace[0.5em]
    \caption{Potential information sources for embedded assessments}
    \label{tab:4}\\
    \endlastfoot
        \textbf{Infrastructure} &
        Developers provide evaluators access to the internal systems and models necessary for the scope of the review &
        \vspace{-0.7em}\begin{itemize}[leftmargin=*, nosep]
\item Monitoring systems and classifiers
\item Chain-of-thought monitors
\item Agent sandboxes and permissions
\item RL training environments
\item Internal state-of-the-art and pre-release models
\item Weight storage and access controls
\item High rate-limit access to models
\item White-box access to intermediate checkpoints if conducting alignment evaluation
\end{itemize}\vspace{-0.2em}\\
        \textbf{Personnel} &
        Developers make relevant staff available for questioning and observation &
        \vspace{-0.7em}\begin{itemize}[leftmargin=*, nosep]
\item Interviews with key members of staff
\item Questionnaires that evaluators can design and key staff fill out
\item Attending key meetings
\item Access to internal messaging channels (e.g. Slack)
\item Step-by-step staff demonstrations of how processes run
\end{itemize}\vspace{-0.2em}\\
        \textbf{Documentation} &
        Developers share the internal records needed to verify their safety claims and spot gaps in internal practices &
        \vspace{-0.7em}\begin{itemize}[leftmargin=*, nosep]
\item Safety assessment results (e.g. monitor robustness assessments, deployment decision records)
\item Evaluation transcripts and elicitation logs
\item Agent logs from internal use
\item Incident reports and logs
\item Safety cases and board minutes
\item Unredacted model and transparency reports
\item Procedural information about how the developer decides a model is safe for internal deployment
\end{itemize}\vspace{-0.2em}\\
\end{xltabular}

No single source of evidence is likely to be sufficient on its own. Most embedded assessments will therefore need to combine infrastructure access, personnel access, and documentation review, allowing evaluators to cross-check findings across multiple sources and build a more complete picture of the risks. The specific range of access to artifacts provided would ultimately depend on the scope and aim of the review. Providing this amount of information to evaluators will create some challenges; however, these can be mitigated, as done in other high-risk industries that use onsite assessments (see \Cref{sec:2.2} for more detail on possible mitigations).

\subsection{How long could embedded assessments last?}\label{sec:3.3}

The duration of an embedded assessment affects how much evaluators can test and learn, but also how burdensome the review is for the developer. There are three broad models: a fixed period, a flexible period that depends on evaluator progress, and a continuous or long-term presence. The appropriate duration will depend on factors such as the scope of the evaluation, evaluator capacity, the rate of change in the developer's systems between assessments, and the pace of progress in AI capabilities.

\textbf{Fixed-duration assessments.} Fixed-duration assessments would run for an agreed period. METR's embedded assessment with Anthropic provides one reference point: one evaluator worked onsite for three weeks \citep{metr_2026c}. The appropriate length would depend on the scope and methods of the assessment, the number of evaluators involved, and, for incident investigations, the severity and complexity of the incident. For example, live red-teaming, interviews, document review, and log analysis may require more time than a narrow review of a single system. However, larger teams may be capable of investigating several areas in parallel, which can increase coverage while reducing the amount of time needed. The review would end on the planned date even if some lines of inquiry remained open, unless an extension were agreed. In METR and Redwood's investigation at OpenAI, OpenAI agreed to two additional onsite visits so the evaluators could clarify questions and address limitations identified in earlier versions of their report \citep{metr_2026d}. This meant the review lasted six days rather than the two initially planned.

The main benefit of fixed-duration assessments is that they make the resource commitment more predictable for both developers and evaluators. Developers can plan staff time, security supervision, workspace, and system access around a known period, while evaluators can plan their other work more easily. As a result, a fixed duration may make developers and evaluators more willing to participate.

However, it may be difficult to determine the appropriate duration before evaluators understand the relevant systems in detail. Consequently, fixed-duration assessments make it more likely that evaluators will run out of time before examining all the information needed to reach confident conclusions. A hard deadline may also discourage evaluators from pursuing unexpected but important findings if doing so would delay completion of the planned assessment. Fixed assessment periods also leave gaps between assessments, meaning they cannot provide lasting assurance because internal systems and practices change quickly as developers frequently deploy new models internally, expand their tool access, change monitoring, and introduce new workflows \citep{kwon_2026}. Together, these limitations could reduce the depth, reliability, and timeliness of the assessment.

\textbf{Flexible-duration assessments.} Flexible-duration assessments would not have a scheduled end date at the outset. Instead, the review would continue until predefined completion criteria were met, such as completing a set of specified tests and documenting a conclusion, or gathering sufficient evidence to address each assessment question to an agreed level of confidence. A maximum duration or a requirement for mutual agreement before extending the review could prevent the commitment from becoming too open-ended.

The main benefit of flexible-duration assessments is that the time available can adapt to the complexity of the systems being assessed and the significance of the findings that emerge as the assessment progresses. This makes it more likely that important issues are investigated thoroughly. It can also give evaluators enough time to debug and rerun evaluations, allowing them to reach more confident conclusions.

However, flexible-duration assessments make costs and staffing less predictable. Without clear stopping rules, there is also a risk that reviews continue after the value of further work becomes low. Developers and evaluators may also disagree about whether sufficient evidence has been collected to bring the review to an end. Once an assessment concludes, there will also be a period before the next review during which material changes or emerging risks will go unexamined.

\textbf{Continuous assessments.} The most comprehensive model would involve evaluators maintaining a continuous or long-term presence within the frontier AI developer \citep{wills_2025}. This would be analogous to resident inspection models used in other high-stakes sectors, such as US meat production \citep{fsis_nd}, nuclear power \citep{nrc_2022}, and banking \citep{occ_nd}.\footnote{In US meat production, animals cannot be slaughtered or dressed for human food unless a federal inspector is continuously present \citep{fsis_nd}. In nuclear power, the NRC stations at least two resident inspectors at each plant \citep{nrc_2022}. In banking, the OCC maintains a continuous, onsite presence at the largest US institutions, those with assets over \$500 billion \citep{occ_nd}.} Anthropic's CEO, Dario Amodei, recently committed to hosting an embedded external review team at Anthropic with ``ongoing access'', providing desks in its offices, company laptops, and permissions comparable to those of internal risk assessment teams \citep{amodei_2026}.

Continuous assessments will typically offer the greatest value because they provide ongoing scrutiny as internal systems and practices evolve. Continuous access would eliminate gaps in oversight between periodic reviews and allow evaluators to identify developments that warrant scrutiny as they occur. Continuous assessments also avoid the periodic set-up costs necessary for a repeated fixed-duration evaluation. Evaluators could also investigate major incidents or system changes quickly without having to establish a new engagement each time. Sustained access would allow them to build knowledge of developer-specific systems, reducing repeated onboarding and information-gathering work. It could also allow evaluators to track remediation through to completion and test whether fixes remain effective as internal systems change.

However, continuous assessments would require more capacity from both evaluators and developers. Evaluators would need to commit staff on an ongoing basis, while developers would need to continuously host evaluators and manage the associated security, intellectual property, and operational risks. Where internal systems change slowly and periodic reviews provide comparable assurance, these costs may be difficult to justify, though we expect such cases to be uncommon. Long-term proximity could also create risks to evaluator independence, although this could be mitigated -- for example, through periodic review of findings by staff from the evaluation organization who are not embedded with the developer.

Overall, given the current rapid pace of progress in frontier AI capabilities, we think continuous assessments are likely already warranted. They would allow evaluators to track changes in internal AI use, investigate emerging concerns promptly, and verify that corrective measures remain effective over time. Where continuous assessments are not yet feasible (e.g. because of limited evaluator capacity), flexible-duration reviews could provide a practical interim approach, particularly for applications of embedded assessments that have not yet been piloted. However, this arrangement would result in far less comprehensive scrutiny than continuous assessments.

\subsection{When could embedded assessments take place?}\label{sec:3.4}

As discussed in \Cref{sec:3.3}, continuous assessments can provide ongoing scrutiny as developers' internal systems and practices evolve. Where continuous assessments are not yet feasible, developers and evaluators need to decide when fixed- or flexible-duration reviews should take place.

We consider four timing approaches: ad hoc, periodic, event-based, and unannounced randomized spot checks. These approaches can also be combined. For example, periodic reviews could provide a regular baseline, while event-based reviews could respond to material changes in risk.

\textbf{Ad hoc (at the developer's discretion).} Ad hoc assessments would take place on a case-by-case basis, with no commitment to repeat them. A developer and evaluator might agree to a review when both think it would be useful, for example after an incident or when examining a new focus area. METR's pilot embedded assessment with Anthropic followed this model: it was conducted as a one-off engagement, with a follow-up only tentatively planned \citep{metr_2026c}.

One benefit of ad hoc evaluations is that they are relatively easy to establish. Because neither party needs to commit to an ongoing program, they can be particularly useful for piloting new approaches and building trust before moving to more regular reviews.

However, ad hoc evaluations provide weak ongoing assurance. Long gaps may arise between engagements, during which time internal systems and practices can change substantially. Future evaluations may also require a completely new agreement, and developers and evaluators may disagree about whether and when another review is needed. Ad hoc reviews may also create selection effects, for example if developers are more willing to commission them when they expect favorable findings, rather than when scrutiny is most needed.

\textbf{Periodic.} Periodic assessments would take place on a fixed schedule (e.g. every three months). The appropriate interval would depend on how quickly the relevant systems are likely to change and how costly the reviews are to conduct.

Periodic assessments provide more regular scrutiny than ad hoc assessments. A fixed schedule reduces reliance on developers or evaluators deciding when another review is needed. Repeated assessments can also test whether earlier weaknesses were fixed and whether safeguards remain effective as systems change. Periodic assessments can also be easier to plan and resource, particularly where access, security, and assessment procedures can be reused. Standardized periodic assessments also allow evaluators to compare findings over time and identify changes in a developer's risk-management performance.

However, fixed schedules may not align with when evaluations are most needed. Frequent reviews also impose ongoing costs on developers and evaluators. In addition, a predictable schedule may give developers time to prepare for the evaluation, for example by curating documents or influencing staff behavior, potentially making it harder to assess how systems operate under ordinary conditions.

\textbf{Event-based.} Event-based assessments would take place when predefined triggers suggest that a review would be particularly valuable. Triggers could include developments that may invalidate previous assessments of internal-use risk, such as major capability increases, increased inference-time compute or tool access, changes to monitoring or agent permissions, serious incidents, or evidence that a safety control has failed \citep{delaney_2026}. Developers and evaluators could agree on these triggers in advance. Developers would then need to notify evaluators when a trigger occurs, after which the developer and evaluator would determine whether a new review is needed and what its scope should be.

The main benefit of event-based evaluations is that they can increase scrutiny when a developer's risk profile has materially changed. This can also help direct scarce evaluator capacity toward times when additional scrutiny is most valuable. Reviews can also be proportionate and tailored to the trigger. For example, a serious incident might justify a targeted investigation of the incident and relevant safeguards rather than a more comprehensive assessment. To illustrate, following the Hugging Face incident, METR and Redwood Research conducted an onsite investigation of the model behavior involved \citep{metr_2026d}. Combined with periodic reviews, event triggers could provide additional assurance between scheduled assessments. Within continuous assessments, these triggers could instead prompt targeted investigations as part of the ongoing engagement.
\newpage

However, event-based evaluations depend on clear triggers and reliable notification from developers that a trigger has occurred. It may be difficult to define triggers precisely enough that developers know when notification is required. Many triggers will only be visible inside the developer, so evaluators will also depend on accurate and timely reporting by the developer. Developers and evaluators may also disagree about whether a trigger warrants a new review or how broad it should be. Some triggers may also prompt scrutiny only after harm has already occurred, for example following an incident or the failure of a safety control.

\textbf{Unannounced (at the evaluator's discretion).} Unannounced assessments would use randomized spot checks. Developers would know a review could occur within an agreed period, but not exactly when. Unannounced onsite inspections are standard practice in other high-stakes domains, including IAEA nuclear safeguards \citep{iaea_nd}, oversight of food and drug manufacturing \citep{fda_2025}, and UK food hygiene enforcement \citep{fsa_2023}.

Unannounced evaluations can provide greater confidence that evaluators are observing normal operating conditions rather than a curated picture of internal systems, documents, and behavior that has been prepared for review. Unannounced reviews could also complement periodic assessments by testing whether conclusions from planned evaluations remain valid between scheduled reviews. The possibility of a spot check could also encourage developers to maintain good practices continuously.

However, truly unannounced reviews may be difficult and costly to organize. Developers may need time to arrange safety and security mitigations for the review, such as vetting, managed devices, supervision, interviews, and system access. Unannounced reviews may also impose additional costs, for example if staff must be made available at short notice or developers need to pause or modify large-scale training runs to accommodate them. Developers may therefore be less willing to accept unannounced reviews voluntarily. Their added value over scheduled assessments may also be limited for assessment targets that are difficult to manipulate in anticipation of a review.

Overall, where continuous assessments are not yet feasible, the timing approaches outlined here can help structure interim reviews. Periodic assessments could provide a regular baseline, supplemented by event-triggered follow-up when risks materially change. Ad hoc reviews may be particularly useful for piloting assessments of new areas and establishing their feasibility before moving toward more sustained scrutiny. However, these time-limited approaches would still leave gaps in oversight and should not be treated as providing assurance equivalent to continuous assessments. They should therefore serve as a bridge toward continuous assessment, rather than a substitute where it is feasible.

\subsection{What terms could govern the assessment?}\label{sec:3.5}

Embedded assessments require developers to give external evaluators deep access to internal infrastructure, personnel, and documentation. This access will usually be governed by an agreement between the evaluator and the developer setting out the terms of the engagement. For example, in cybersecurity, engagement letters and rules of engagement between the security tester and the company being evaluated are agreed in advance \citep{nist_2008}.

These terms should give evaluators sufficient access and independence to conduct a credible review while limiting security, intellectual property, and legal risks. The agreement should address four areas in particular: access rights; security and information handling; evaluator independence and conflicts of interest; and legal protections and safe harbor.

\textbf{Access rights.} Evaluators would agree on the rights to access the systems, information, and staff needed to assess the relevant safety claims before the review begins. The agreement would also establish a process through which evaluators can request and obtain additional access where this becomes necessary to conduct the assessment, for example so that they can pursue unexpected findings.

The benefit of setting out these access rights in advance is that they give evaluators greater assurance that they will be able to conduct the assessment with access to all necessary information. These rights should establish parity of access with the developer's senior staff conducting similar risk assessments or senior alignment researchers, as appropriate. Clear access rights reduce the risk that developers restrict access once a review is underway or prevent evaluators from following up on concerning findings. At the same time, it may be difficult to anticipate every form of access an evaluator will need. A mechanism for requesting additional access therefore gives evaluators flexibility to investigate unexpected issues without requiring every possible need to be specified beforehand.

However, this does create costs. Developers may find it harder to predict and manage the security burden of an assessment, and reviews could be delayed while additional requests are considered. Conversely, granting new access too quickly may expose sensitive systems or information without adequate safeguards. The parties may also disagree over whether a particular request is genuinely necessary for the assessment. Nevertheless, clear access rights benefit both parties because they help ensure evaluators receive sufficient access to reach confident conclusions and allow developers to specify what safeguards are required to reduce the resulting risks.

\textbf{Security and information-handling requirements.} Security requirements can limit access to what is necessary and help detect and respond to misuse, without unduly constraining legitimate evaluation activity. These could include company-managed devices, least-privilege access, session limits, two-factor authentication, personnel vetting, escorted or mediated access, activity logging, anomaly detection, incident-response procedures, and access-revocation terms \citep{nist_2020}. Information-handling requirements can reduce the risk that sensitive information leaves the company or is retained unnecessarily. These could include NDAs, restrictions on offsite copies, and rules for storing, retaining, deleting, reviewing, or escrowing evaluator notes and other sensitive material \citep{nist_2008, nist_2020}.

These safeguards offer several benefits. They may make developers more comfortable providing evaluators with broad access to sensitive systems and information. For example, restricting offsite copies could support a relatively permissive approach to access within the company. Logging, anomaly detection, and pre-agreed response procedures could also help identify both scope creep, where evaluators access systems or information beyond the agreed review, and misuse of otherwise authorized access.

However, despite their importance, these controls do have some downsides, especially when used inappropriately. Measures such as escorted access and personnel vetting may increase costs and slow the review. More importantly, overly restrictive terms could prevent evaluators from pursuing unexpected findings. Excessive control by the developer over evaluator access, notes, or records could also undermine the independence or perceived credibility of the assessment. It is therefore important to strike a balance between protecting sensitive systems and information and preserving evaluators' ability to conduct a thorough, independent assessment.

\textbf{Evaluator independence.} The terms could include safeguards to protect evaluator independence. Without them, evaluators may face incentives to soften findings if they depend on the developer for future work or develop close relationships with its staff. Several measures could help mitigate these risks. Evaluators could be required to rotate between successive reviews or observe a cooling-off period before joining a developer they assessed or one of its competitors, as is required in nuclear power, banking supervision, and auditing (\citealp{nrc_2022}; \href{https://www.law.cornell.edu/uscode/text/12/1820}{12 U.S.C. § 1820(k)}; \href{https://www.law.cornell.edu/uscode/text/15/78j-1}{15 U.S.C. § 78j-1}). Evaluators could additionally be required to disclose relevant financial, professional, or personal conflicts, as financial auditing firms are required to do \citep{pcaob_nd}.\footnote{Before accepting an engagement and annually thereafter, registered audit firms must describe in writing all relationships that may reasonably be thought to bear on their independence, discuss their potential effects, and affirm in writing that the firm is independent \citep[Rule 3526]{pcaob_nd}. SEC rules define the financial, employment, and business relationships that impair independence (\href{https://www.ecfr.gov/current/title-17/chapter-II/part-210/section-210.2-01}{17 C.F.R. § 210.2-01}).}

These measures address different threats to independence. Cooling-off periods can reduce incentives to soften findings in anticipation of future employment. Rotation can reduce familiarity bias and the risk of evaluator capture across repeated engagements. Conflict-of-interest disclosures can make other potential sources of bias more visible and easier to manage.

However, each measure also involves trade-offs. Rotation reduces continuity and requires new evaluators to become familiar with developer-specific systems. It also relies on evaluators having enough capacity for this to be feasible.\footnote{In cases where this is not possible other mitigations can be used, for example by ensuring periodic review of findings by staff from the evaluation organization who are not embedded with the developer.} Cooling-off periods restrict career options and could make evaluator recruitment more difficult. More broadly, a blanket exclusion of anyone with prior connections to a frontier AI developer would likely be both impractical and undesirable given the small pool of people with relevant expertise. This means proportionate disclosure and management of conflicts are likely to be particularly important measures.

\textbf{Legal protections.} Evaluators could have appropriate protection from legal liability for testing conducted within the agreed scope and for reporting findings through the agreed process. One option is to include an explicit legal safe harbor in the evaluation agreement \citep{longpre_2024}.\footnote{By ``safe harbor'', we mean a contractual commitment that the developer will not pursue legal action against evaluators for good-faith activities carried out within the agreed scope. Similar protections are used in cybersecurity vulnerability-disclosure programs. For example, CISA recommends that organizations treat good-faith research conducted under their policy as authorized and commit not to pursue legal action against the researcher \citep{cisa_2020}. \citet{longpre_2024} propose legal and technical safe harbors to protect good-faith AI evaluators and red teamers from account suspension or legal reprisal.} The agreement would need to specify which activities are protected and the circumstances in which those protections would cease to apply. The agreement could also establish procedures for handling legally compelled disclosure, including notifying the developer where legally permitted and seeking protective orders or confidential treatment for sensitive information where available. The agreement could also preserve any applicable whistleblowing rights.

These protections can reduce incentives for evaluators to narrow their testing or soften their conclusions because of perceived or actual legal risk \citep{anderljung_2023}. A carefully designed safe harbor can also reinforce limits on evaluator conduct by making protection conditional on evaluators following agreed methods and remaining within the authorized scope. Clear procedures for compelled disclosure can help both parties coordinate their response to legal demands while seeking to protect sensitive information. Explicitly preserving whistleblowing rights can also reassure evaluators that the agreement does not prevent them from making protected disclosures, including where agreed escalation routes are absent or ineffective.

The main challenge is defining these boundaries precisely. A safe harbor could protect legitimate evaluation activity without shielding conduct that falls outside the agreed terms. The parties may also disagree about how to reconcile confidentiality requirements with disclosure obligations and applicable whistleblowing rights. For example, developers may seek notification and review procedures to protect sensitive information, but overly restrictive requirements could delay disclosures or discourage evaluators from raising concerns. Negotiating these provisions may also make evaluation agreements more complex and costly to establish for both evaluators and developers.

Overall, the agreement governing an embedded assessment should set clear terms for evaluator access, security and information handling, evaluator independence, and legal protections. Agreeing these terms in advance can help developers manage security and legal risks while giving evaluators sufficient access and autonomy to conduct a rigorous and credible assessment. At a minimum, evaluators should be meaningfully independent: they should retain full editorial control, disclose and manage relevant conflicts of interest, and avoid significant commercial ties or outcome-contingent payments (or other rewards) from the developers they assess \citep{aef_2026}.

\subsection{How could findings be publicly disclosed?}\label{sec:3.6}

This section focuses only on public disclosure. We assume that evaluators would share their full private findings with relevant staff at the frontier AI developer before publication, as in pilot embedded assessments conducted by METR \citep{metr_2026d, metr_2026c}. We discuss escalation of findings to other types of stakeholders in \Cref{sec:3.7}.

Public disclosure can vary along a spectrum, from disclosing nothing about the review to providing detailed or ongoing public reporting. We distinguish four broad approaches: no public information about the review, a statement that the review occurred, a summary report, and a detailed report.

\textbf{No information about the review.} Under this approach, no information about the assessment would be disclosed publicly, including that it took place or what it found. Findings would still be reported privately to the developer and potentially other actors.\footnote{In practice, findings could be shared company-wide or only with designated individuals. We do not examine this choice in detail here, although \Cref{sec:3.7} discusses escalation procedures, including escalation to a senior employee responsible for overseeing the developer's response, an independent oversight body, or a relevant government body.}

This approach minimizes the risks associated with public disclosure. It reduces the risk of exposing vulnerabilities, proprietary information, or other sensitive material. It also limits the reputational and legal risks associated with publishing findings. It may therefore make developers more willing to host embedded assessments.

However, this approach provides no public transparency or accountability. External audiences would be unable to assess the scope, quality, or conclusions of the review, and developers would face no public pressure to address serious concerns. Although the assessment could still inform improvements within the developer's organization, its contribution to external scrutiny would be considerably limited.

\textbf{A statement that the review occurred.} Under this approach, evaluators would publicly disclose that a review took place, including the developer involved, the purpose of the assessment, and its dates, but would not publish any findings. Findings would still be reported privately to the developer and potentially other actors.\footnote{See footnote 31.}

This approach provides some transparency while limiting the risks associated with publishing the findings of the assessment. Publicly confirming that an independent review took place can provide external assurance to a limited extent, while avoiding the potential security, intellectual property, legal, and reputational risks associated with publishing findings. This may make developers more willing to participate while preventing reviews from taking place entirely outside public view.

However, a statement that a review occurred provides very limited accountability. It gives external audiences no information about what the evaluation actually found and provides no basis for assessing whether the developer adequately addressed any issues identified by evaluators. It could therefore create a false sense of security if participation were interpreted as evidence that the developer's practices were adequate.

\textbf{A summary report.} Under this approach, evaluators would publish a summary of the assessment's scope, high-level findings, the access they received, key limitations and uncertainties, and their recommendations, without presenting the supporting evidence in detail. A standard rating system could help readers compare results across developers and over time, where assessment scopes and methods are sufficiently comparable. The developer could also provide an accompanying response identifying which recommendations it intends to adopt, who is responsible for implementing them, and the expected timeline.

Developers could have a time-limited opportunity to identify factual errors and request redactions of sensitive information, while evaluators retain editorial control over the report.\footnote{Evaluators have retained editorial control over public reports in previous assessments. For example, in METR's Frontier Risk Report, Anthropic, Google, Meta, and OpenAI provided access to nonpublic information while allowing METR to publish its report without requiring approval from the frontier AI developers.} These redaction limits resemble those proposed by Anthropic's CEO, Dario Amodei, under which the developer could redact security-sensitive, legally privileged, commercially sensitive, or third-party confidential information, but not findings simply because they are unfavorable \citep{amodei_2026}. Evaluators could indicate whether any redactions are material to their conclusions. Unresolved disagreements could also be published alongside the findings. Separately, the agreement could also include an exit clause allowing the developer to block publication of findings, while still requiring public disclosure of the engagement's purpose and dates.

A key benefit is that it provides some public information about the findings while potentially requiring less work to prepare and review for sensitive information than a more detailed report. Disclosing the access evaluators received and the limitations they encountered can help readers judge how much confidence to place in the findings. Redaction rights and, more substantially, an exit clause could lower barriers to participation. Publishing the developer's planned follow-up actions could also increase accountability for addressing the issues identified.

However, the limited detail would prevent external audiences from being able to assess the evidence underlying the evaluator's conclusions. Developers and evaluators may also disagree about which findings belong in the summary, potentially delaying publication or excluding relevant context. For example, there may be pressure to omit a near-miss incident because it caused no actual harm, even though it revealed an important weakness. An exit clause creates a further limitation: although it may encourage participation, it allows the developer to prevent adverse findings from reaching the public, reducing disclosure to a statement that the review occurred.

\textbf{A detailed report.} Under this approach, evaluators would publish an account that explains not only their conclusions but also the evidence and reasoning supporting them. In addition to the information included in a summary report, it would describe the assessment methods, specific findings and supporting evidence, and unresolved questions. The report could also summarize the key terms of engagement and explain any constraints they placed on evaluators' access, independence, or ability to report findings.

The same publication and follow-up arrangements described for summary reports could apply, including a time-limited opportunity for the developer to identify factual errors, narrowly defined redaction rights, evaluator editorial control, and potentially an exit clause. In either reporting format, evaluators could state whether redactions are material to their conclusions. As with summary reports, the developer could also publish a response identifying which recommendations it intends to adopt, who is responsible for implementing them, and the expected timeline, followed by periodic progress updates. Evaluators could also recommend remediation periods and subsequently assess whether the developer has adequately addressed the concerns.

The additional detail provides a stronger basis for external scrutiny. Readers can examine the evidence and reasoning behind the findings, identify gaps, and distinguish well-supported conclusions from more tentative judgments. Although both reporting formats can support accountability through public recommendations and developer responses, detailed reports give readers more information with which to judge whether the proposed actions adequately address the concerns. They may also provide more useful lessons for other developers and evaluators by explaining how weaknesses were identified and assessed.

However, compared with summary reports, detailed reports may require more work to prepare and review, and create greater risks of exposing vulnerabilities, proprietary information, or sensitive details about internal systems. The same narrowly defined redaction rights could help manage these risks, but applying them to a larger body of supporting evidence may require more extensive discussions and delay publication. Redactions could also remove information that readers need to scrutinize particular conclusions. These additional demands and disclosure risks may make developers less willing to participate voluntarily. As with summary reports, an exit clause could lower barriers to participation, but at the cost of allowing the developer to prevent adverse findings from being published altogether.

Existing reporting already provides useful precedents for detailed disclosure. METR and Redwood Research's OpenAI investigation, for example, describes its scope, methods, evidence, access limitations, and the significance of redactions \citep{metr_2026d}. Nevertheless, the publication arrangements allowed OpenAI to redact any nonpublic information beyond the agreed description of the engagement's scope and terms. METR's broader Frontier Risk Report, which incorporates findings from the Anthropic embedded assessment, similarly notes that some supporting context was removed without a specific public indication of each omission \citep{metr_2026b}. These examples suggest that the priority is not simply longer reports, but clearer commitments about what will be disclosed and guarantees that developers cannot block publication of evaluators' findings.

We recommend building on these precedents by having evaluators publish detailed public reports at least quarterly. Developers should have narrowly defined redaction rights, while evaluators should retain editorial control and explain whether any redactions affect their conclusions. Developers should also publish which recommendations they intend to adopt together with their implementation plans (or an explanation for why they will not adopt key recommendations), and report periodic progress updates. Together, these arrangements would allow external stakeholders to scrutinize both the assessment's findings and the developer's response, while protecting sensitive information.

\subsection{What escalation procedures could be put in place?}\label{sec:3.7}

Embedded assessments may uncover serious weaknesses that require more than simply communicating the findings publicly. A formal escalation process can help ensure that serious or unresolved concerns reach people with sufficient authority to act. Broadly, options range from no formal process to mandatory escalation to a senior employee at the developer who is responsible for safety and oversees its response. More extensive arrangements could allow further escalation to an independent oversight body or, in particularly serious cases, a relevant government body.

\textbf{No formal escalation procedure.} Under this approach, evaluators would report their findings to the developer, but there would be no agreed process for escalating or addressing serious concerns. This appears to be the approach taken in METR's pilot with Anthropic: though Anthropic acted quickly on some serious findings,\footnote{For example, Anthropic fixed a bug that left subagent calls unmonitored less than 24 hours after METR identified it \citep{metr_2026b}.} METR's public report does not describe a formal escalation process that would have applied had Anthropic declined to remediate the identified vulnerabilities.

One advantage of having no formal escalation procedure is that it minimizes additional governance requirements for participating developers. Instead, it allows developers to respond through their existing decision-making and incident-response processes. This may make voluntary participation easier by avoiding commitments about how findings must be handled.

However, this approach provides little assurance that significant findings will reach senior decision makers or be acted on. Evaluators would also have no formal route to escalate concerns if they believed the developer's response was inadequate.

\textbf{Escalation to a designated senior employee responsible for safety.} Under this approach, evaluators would be required to share the findings of their review with a designated senior employee responsible for safety within the developer's organization, such as Anthropic's Responsible Scaling Officer. That person would be formally accountable for reviewing the findings and recommendations and overseeing the developer's response.

The main benefit of this approach is that it helps ensure that important findings reach someone with clear responsibility for company safety. Requiring escalation to a designated employee may reduce the risk that concerns are handled informally or remain within the team whose work is being evaluated. It is also relatively lightweight and could fit within existing governance structures.

However, escalation to a single employee still has limitations. For example, the designated employee may have incentives to downplay difficult findings or may lack sufficient authority with the board or senior leadership to ensure that concerning findings are acted on, especially in cases where this would impose significant costs on the developer. Evaluators would also have limited recourse if they believed the response remained inadequate.

\textbf{Escalation to an independent oversight body.} Under this approach, evaluators would initially report findings through the developer's existing governance structures (e.g. a senior employee responsible for safety). Significant or unresolved concerns could then be escalated to an independent oversight body, such as the board.\footnote{Note that evaluators may not know whether the developer has failed to address any concerns until a subsequent review if it is not continuous.}

An important benefit of this approach is that it provides a second layer of oversight if management or the designated safety lead does not, or cannot, respond adequately. It also increases the chance that serious concerns reach people with sufficient authority to require action.

However, this approach depends on having clear escalation thresholds and an effective oversight body.\footnote{Which criteria should trigger escalation is an important design question, but one we do not examine here.} It may be difficult to define when a concern is sufficiently serious or unresolved to justify escalation. Its value also depends on the oversight body being sufficiently independent and having the authority to compel action.

\textbf{Escalation to a government body.} Particularly serious or unresolved findings could be escalated outside the developer to a relevant government body. In such cases, the evaluator could write a tailored, largely unredacted report to a government body describing their findings and concerns. The assessment agreement could define in advance the threshold for external escalation, the body that would receive the escalation, and the type of information that would still need to be redacted. Escalation could be limited to particularly serious cases, such as potential legal violations or imminent threats to safety.

The main benefit of government escalation is that it provides an external route for investigating and addressing serious concerns when internal governance mechanisms fail. Government bodies may have powers to investigate, require remediation, or impose other measures. This could help ensure that concerns with significant public-safety or national-security implications are investigated and addressed even where the developer disputes the findings or declines to act despite other escalation measures.

However, external escalation could increase legal, regulatory, and reputational risks for developers. The possibility that evaluators could report findings directly to government may also make developers less willing to participate voluntarily.

Overall, a tiered escalation process may provide the clearest route for ensuring that serious findings are acted on. Significant concerns could first be reported to a designated senior employee responsible for safety, with particularly serious or unresolved issues escalated to an independent oversight body and, in exceptional cases, a relevant government body. This would preserve an opportunity for the developer to respond internally while providing additional routes for escalation where that response is inadequate. However, external escalation may be difficult to establish while embedded assessments remain voluntary.

Alongside these formal escalation procedures, individual evaluators should have appropriate protections for whistleblowing. They should be able to report reasonably held concerns about serious safety risks or wrongdoing to relevant authorities without retaliation from either the developer or their evaluation organization. This is particularly important where no external escalation route has been agreed or where the agreed process fails to address the concerns. Assessment agreements should preserve any applicable whistleblowing rights and include confidentiality exceptions and protections against retaliation to support such reporting (see \Cref{sec:3.5}).\footnote{Whistleblowing protections may create additional challenges, such as bringing added liability risk for developers. It may also be difficult to set practical boundaries on which information or events evaluators could whistleblow on.}

\section{Recommendations}\label{sec:4}

The design choices discussed in \Cref{sec:3} can be combined in different ways. Below, we set out recommendations for developers and evaluators to act on now, summarized in \Cref{tab:5}. These are intended as a starting point: we hope they will help establish good practices for embedded assessments and provide a foundation on which to build further. We then outline some of the highest-priority ways to build on these recommendations, though these are not exhaustive and we recognize that these priorities may need to change as the AI landscape evolves.

\renewcommand{\arraystretch}{1.2}
\begin{xltabular}{\linewidth}{%
        >{\small\raggedright\arraybackslash}p{\dimexpr0.2333\linewidth-2\tabcolsep\relax}%
        >{\small\raggedright\arraybackslash}p{\dimexpr0.3833\linewidth-2\tabcolsep\relax}%
        >{\small\raggedright\arraybackslash}X}
    \toprule
        \textbf{Design choice} &
        \textbf{Recommendation} &
        \textbf{Rationale}\\
    \midrule
    \endfirsthead
    \toprule
        \textbf{Design choice} &
        \textbf{Recommendation} &
        \textbf{Rationale}\\
    \midrule
    \endhead
    \bottomrule
    \endfoot
    \bottomrule
    \addlinespace[0.5em]
    \caption{Recommended starting point for embedded assessments}
    \label{tab:5}\\
    \endlastfoot
        \textbf{Assessment target} &
        At a minimum, embedded assessments should cover three areas central to managing risks from internal AI use: internal agent monitoring, internal agent security controls and permissions, and model alignment. &
        These areas are high priority, with recent incidents highlighting gaps in monitoring and failures of containment. They are also tractable for embedded assessments now, and depend on internal systems and practices that are difficult to assess through existing forms of external oversight.\\
        \textbf{Information sources} &
        Evaluators should have parity of access with internal staff conducting similar risk assessments or senior alignment researchers, as appropriate. This should include relevant infrastructure (e.g. monitoring systems and agent sandboxes), personnel (e.g. interviews and internal communication channels), and documentation (e.g. agent logs and incident reports), scoped to the assessment targets. Access to relevant information should be granted by default. Requests for additional information should be granted unless the developer has a specific reason for refusal, which it should share with the evaluator. &
        No single source of information is likely to provide enough evidence to adequately assess risks. Combining sources allows evaluators to test systems directly, understand how they operate in practice, and verify claims against internal records.\footnotemark{} Access by default, including additional information requested during the assessment, helps evaluators pursue unexpected findings and obtain the evidence needed to reach independent conclusions.\\
        \textbf{Timing and duration} &
        Embedded assessments should operate continuously, with serious incidents or material changes prompting targeted investigations. Where continuous assessment is not yet feasible (e.g. when piloting a new assessment area or where evaluator capacity is limited), flexible-duration reviews conducted periodically with event-triggered follow-ups can provide an interim approach. &
        Findings can quickly become outdated as internal models, safeguards, and practices change. Continuous assessments support sustained scrutiny, timely investigation of emerging concerns, and follow-up on recommended actions.\footnotemark{} Targeted investigations following serious incidents can identify the causes and inform mitigations to reduce the risk of recurrence. Interim arrangements provide less comprehensive scrutiny and should serve as a bridge toward continuous assessment only where needed.\\
        \textbf{Terms} &
        Developers and evaluators should agree on terms before the assessment begins, covering (1) access rights (parity of access and a process for requesting additional access), (2) security and information-handling requirements (safeguards such as company-managed devices, restrictions on offsite copies, and rules for storing and retaining sensitive information), (3) evaluator independence (evaluators should retain editorial control over their reports, disclose and manage conflicts of interest, and avoid significant commercial ties to the developer or payments tied to their findings), and (4) legal protections (protection for testing within the agreed scope and reporting findings through the agreed process, for example through a contractual safe harbor). &
        Pre-agreed terms help ensure evaluators receive sufficient access and independence to conduct a credible assessment and report their findings, while allowing developers to manage security, intellectual property, and legal risks. Sufficiently expansive formal access rights provide assurance that evaluators will be able to access the information they need for the assessment, while procedures for obtaining additional access preserve flexibility as new questions arise.\\
        \textbf{Public disclosure} &
        Evaluators should publish detailed reports at least quarterly, setting out the assessment's scope, findings and supporting evidence, key limitations and uncertainties, and any recommendations that have not yet been addressed. Developers should have a narrowly defined right to redact sensitive information, but evaluators should retain editorial control and state whether any redactions are material to their conclusions. Developers should publish a separate report identifying which recommendations they intend to adopt, an implementation plan and timeline, and reasons for declining any recommendations. They should also provide periodic updates on their progress. &
        Detailed and regular reporting allows external stakeholders to scrutinize findings and their evidential basis.\footnotemark{} Narrow redaction rights protect sensitive information without allowing unfavorable findings to be suppressed. Developer responses, implementation plans, and progress updates strengthen accountability for acting on recommendations.\\
        \textbf{Escalation} &
        Findings and recommendations should be shared with a designated senior employee responsible for safety (e.g. Anthropic's Responsible Scaling Officer). This person should be formally accountable for reviewing them and overseeing the developer's response. Serious or unresolved concerns should be escalated to the developer's board or another designated oversight body. &
        Assigning responsibility to a named senior employee increases the likelihood that findings are acted on. Further escalation provides an additional route where concerns are particularly serious or the initial response is inadequate.\\
\end{xltabular}
\addtocounter{footnote}{-3}
\stepcounter{footnote}\footnotetext[\value{footnote}]{Anthropic's CEO, Dario Amodei, similarly proposes giving embedded evaluators access to tools, permissions, and employees broadly similar to those of internal risk-assessment teams \citep{amodei_2026}.}
\stepcounter{footnote}\footnotetext[\value{footnote}]{Anthropic's CEO, Dario Amodei, has already committed to setting up embedded evaluators at Anthropic on a continuous basis \citep{amodei_2026}.}
\stepcounter{footnote}\footnotetext[\value{footnote}]{A quarterly reporting cadence also aligns with the default schedule in California's SB 53 for reporting assessments of catastrophic risk from internal AI use.}

\textbf{Recommended starting point for embedded assessments.} We recommend that assessments cover, at a minimum, the three areas of internal AI use examined in this paper. Assessments of internal agent monitoring should examine whether consequential activity is adequately covered, whether monitors reliably detect harmful behavior, and whether resulting flags are escalated and reviewed appropriately. This should include assessing whether changes in the monitorability of model reasoning undermine detection. Assessments of internal agent security controls and permissions should examine whether agents are appropriately restricted and whether they can bypass those restrictions. Assessments of model alignment should involve evaluations to examine model behavior across the lifecycle. They should also assess whether current alignment techniques are likely to remain effective for more capable near-future models.

These assessments should be continuous. Ongoing access is needed so that evaluators can track changes in internal systems and practices, investigate emerging concerns, and verify whether corrective measures remain effective. Serious incidents or other material changes should prompt more focused investigations into the relevant risks and safeguards. For other applications, such as assessing the effectiveness of security mitigations protecting sensitive assets (e.g. model weights), embedded assessments may initially need to be piloted for an agreed period. However, they should not be treated as offering equivalent assurance, and as with assessments of internal AI use, they should then move toward continuous assessment once pilots have established their feasibility.

Meaningful scrutiny also requires sufficient access and independence. Evaluators should have parity of access with the developer's staff conducting similar risk assessments or senior alignment researchers, including access to relevant infrastructure, personnel, and documentation. Pre-agreed terms should cover access rights, security and information-handling requirements, evaluator independence, and legal protections. They should establish sufficient access and allow it to be adjusted as new questions emerge. Requests for additional relevant information should be granted unless developers can offer specific reasons for any denial, which should be shared with the evaluator. Security and information-handling requirements should support this access rather than unduly restrict evaluators' ability to investigate important questions.

Evaluators should publish detailed public reports at least quarterly, setting out the assessment's scope, findings and supporting evidence, key limitations and uncertainties, and outstanding recommendations. They should retain editorial control and state whether any redactions are material to their conclusions. Before publication, developers should have a narrow right to redact sensitive information, and should not be able to redact findings simply because they are unfavorable. A designated senior employee responsible for safety should be formally accountable for overseeing the developer's response to the findings and recommendations. Serious or unresolved concerns should be escalated to the developer's board or another oversight body. Developers should also publish a response identifying which recommendations they intend to adopt, an implementation plan and timeline, and reasons for declining any recommendations, followed by periodic progress updates.

\textbf{High-priority extensions for near-term embedded assessments.} The three areas of internal AI use identified above are a minimum, not a limit on the scope of embedded assessments. Developers and evaluators should also pursue opportunities to broaden assessments and strengthen their ability to provide third-party scrutiny of the risks posed by frontier AI. The extensions below can be pursued alongside the recommendations above, rather than only once those recommendations are fully implemented.

First, embedded assessments should cover a broader range of activities, including security mitigations protecting sensitive assets such as model weights and compliance with potential future pacing commitments (see the Appendix). Broader scrutiny could provide a more complete picture of how developers manage risks across model development and deployment. It could also help identify gaps between technical safeguards and organizational processes, inform improvements to company practices, and provide stronger assurance that risks are being managed effectively.

Second, embedded assessments should support stronger external accountability. Particularly serious or unresolved concerns should be escalated to an appropriate government body in cases where other governance mechanisms, including escalation to the board or another oversight body, have failed to address them. Assessment agreements should specify the conditions for external escalation, the body that would receive the findings, and procedures for sharing sensitive evidence. This would provide a backstop where internal escalation does not lead to an adequate response to significant safety concerns.

Third, evaluator capacity and expertise should grow alongside the scope of assessments. Evaluator teams should collectively have the expertise needed to assess a broader range of technically demanding areas, including by involving multiple evaluation organizations where necessary. This would help broaden coverage without sacrificing the depth of scrutiny within each area.

\section{Conclusion}\label{sec:5}

Embedded assessments could fill an important gap in frontier AI oversight. They could enable deeper and more flexible scrutiny of risks that depend on a developer's internal systems and practices under stronger security controls. This paper has examined the benefits and challenges of embedded assessments and set out options for seven key design questions. We hope this framework helps developers and evaluators establish credible assessments while managing the security, intellectual property, legal, and capacity challenges involved.

We recommend that frontier AI developers begin hosting embedded assessments now, examining, at a minimum, internal agent monitoring, internal agent security controls and permissions, and model alignment. These assessments should operate continuously, allowing evaluators to track changes over time, investigate emerging concerns, and follow up on corrective measures. Meaningful scrutiny requires sufficient access, evaluator independence, and accountability for responding to findings. Evaluators should have parity of access with the developer's staff conducting similar risk assessments or senior alignment researchers, as appropriate. Pre-agreed terms should protect this access and evaluators' independence while managing the associated risks. Evaluators should publish detailed public reports at least quarterly, retaining editorial control and allowing only narrowly defined redactions of sensitive information. Developers should publish their planned responses and progress updates, with clear responsibility for reviewing findings and escalation routes for serious or unresolved concerns.

These recommendations are a starting point, not a limit on the scope or ambition of embedded assessments. Broader assessments could provide a more complete picture of how developers manage risks across model development and deployment, including how they protect sensitive assets such as model weights. Stronger external accountability, including escalation to relevant government bodies where other governance mechanisms fail, could help ensure that serious concerns are addressed. Developers and evaluators should pursue these priorities alongside implementing the recommendations above. Embedded assessments will also require greater evaluator capacity, without compromising technical expertise or independence. Teams should bring together the expertise and differing viewpoints needed for the areas they assess, including through the involvement of multiple evaluation organizations where appropriate. Further work could develop contractual templates, specify more detailed terms of engagement, and examine appropriate evaluation-team composition.

\section*{Acknowledgements}
\addcontentsline{toc}{section}{Acknowledgements}

We are grateful for valuable comments and feedback from Aidan Homewood, Alex Meinke, Annika Hallensleben, Charlotte Stix, David Rein, Jacob Davies, Marius Hobbhahn, Patricia Paskov, and Theodore Ehrenborg (in alphabetical order).

\newpage
\section*{Appendix: Other potential applications for embedded assessments}
\addcontentsline{toc}{section}{Appendix: Other potential applications for embedded assessments}

Below are some other potential applications for embedded assessments, in addition to those we focus on in this paper. There are likely many other applications, and the table is not meant to be exhaustive.

\renewcommand{\theHtable}{uncaptioned.appendix}
\renewcommand{\arraystretch}{1.2}
\begin{xltabular}{\linewidth}{%
        >{\small\raggedright\arraybackslash}p{\dimexpr0.2333\linewidth-2\tabcolsep\relax}%
        >{\small\raggedright\arraybackslash}p{\dimexpr0.3833\linewidth-2\tabcolsep\relax}%
        >{\small\raggedright\arraybackslash}X}
    \toprule
        \textbf{Application} &
        \textbf{Description} &
        \textbf{Rationale}\\
    \midrule
    \endfirsthead
    \toprule
        \textbf{Application} &
        \textbf{Description} &
        \textbf{Rationale}\\
    \midrule
    \endhead
    \bottomrule
    \endfoot
    \bottomrule
    \endlastfoot
        \textbf{Security mitigations} &
        Embedded evaluators could assess controls protecting sensitive assets, such as model weights. &
        Verifying the effectiveness of these controls may require red-teaming live security infrastructure (either as an external attacker or a misaligned internal model), alongside reviewing internal security documentation. This is impossible through APIs or externally shared artifacts alone.\\
        \textbf{Pacing commitments or agreements} &
        Embedded evaluators could verify developers' adherence to potential future pacing commitments, such as capability-based checkpoints or limits on training and inference compute, model deployment, or the internal use of AI to accelerate AI development. &
        Pacing commitments may impose highly specific technical constraints on model architecture, training, inference, and deployment, making assessment of compliance dependent on access to relevant internal systems and records, as well as the technical staff responsible for implementing the commitments.\\
        \textbf{Model capabilities} &
        Embedded evaluators could assess the capabilities and risk-relevant propensities of internal or pre-release models. &
        Some developers may be more willing to grant deeper access to models and sensitive training artifacts to evaluators who are onsite. Onsite arrangements could also provide stronger security controls for assessing dangerous capabilities, such as those that could facilitate cyberattacks or biological misuse, than providing remote API access to the model.\\
        \textbf{Training practices} &
        Embedded evaluators could assess the safety of developers' broader training practices (e.g. training-data integrity and reinforcement learning environment design). This would involve closer integration with the entire capabilities team, which goes beyond what we recommend in \Cref{sec:3.1.3}. &
        Assessing these practices requires access to training pipelines and observation of the staff who design and run them.\\
        \textbf{Governance} &
        Embedded evaluators could assess developers' broader safety and governance processes, including whether they are adequate and consistent with their published safety frameworks. &
        Embedded access could allow evaluators to observe how decisions are made, escalated, and documented in practice, and identify gaps between written policies and their implementation that may be difficult to assess remotely.\\
        \textbf{Regulatory compliance} &
        Evaluators could assess whether developers meet applicable regulatory requirements relating to risk management. &
        Relevant evidence may be distributed across internal systems, records, and teams. Embedded access could help evaluators cross-check this evidence and assess whether required practices are implemented, rather than relying solely on externally shared documentation.\\
\end{xltabular}
\addtocounter{table}{-1}
\renewcommand{\theHtable}{\arabic{table}}

\newpage
\bibliographystyle{apacite}
\bibliography{ms}

\end{document}